\documentclass[twocolumn,showpacs,preprintnumbers,amsmath,amssymb]{revtex4-1}
\usepackage{graphicx,color}
\usepackage{dcolumn}
\usepackage{bm}
\usepackage{epstopdf}
\usepackage{tabularx}

\begin{document}

\preprint{APS/123-QED}

\title{Spin-lattice interaction
  parameters from first principles: theory and implementation}

\author{Sergiy Mankovsky}
\affiliation{%
Department of Chemistry/Phys.\ Chemistry, LMU Munich,
Butenandtstrasse 11, D-81377 Munich, Germany 
}%
\author{Hannah Lange}
\affiliation{%
Department of Chemistry/Phys.\ Chemistry, LMU Munich,
Butenandtstrasse 11, D-81377 Munich, Germany 
}%
\author{Svitlana Polesya}
\affiliation{%
Department of Chemistry/Phys.\ Chemistry, LMU Munich,
Butenandtstrasse 11, D-81377 Munich, Germany 
}%
\author{Hubert Ebert}
\affiliation{%
Department of Chemistry/Phys.\ Chemistry, LMU Munich,
Butenandtstrasse 11, D-81377 Munich, Germany 
}%

\newcommand{\BLUE}{\bf \color{blue} }
\newcommand{\RED}{\bf \color{black} }
\newcommand{\DONE}[1]{ \marginpar{\bf\large DONE} {\bf \em #1}  \marginpar{\bf\large DONE}}

\newcommand{\DISC}{ \marginpar{\bf\em\large DIS\-CUS\-SION} }  

\date{\today}

\begin{abstract}
  A scheme is presented to calculate on a first-principles level
  the spin-lattice coupling (SLC) parameters needed to perform 
  combined molecular-spin dynamics (MSD) simulations.
By treating  changes to the spin configuration 
and atomic positions on the same level,
closed expressions for the atomic SLC parameters
could be derived  in a coherent way up to any order.
The properties of the SLC parameters are discussed considering
separately the symmetric and antisymmetric parts of the SLC tensor. 
The changes due to atomic displacements of the spin-spin exchange coupling (SSC)
parameters estimated using the SLC parameters are compared with
the SSC parameters calculated for an embedded cluster with the
central atom displaced, demonstrating good agreement of these results. Moreover,
this allows to study the impact of different SLC contributions, linear and
quadratic with respect to displacements, on the properties of the
modified SSC parameters. 
In addition, we represent an approach to calculate the site-diagonal SLC
parameters characterizing local magnetic anisotropy induced by a lattice
distortion, which is a counterpart of the approach based on magnetic
torque used for the investigations of magneto-crystalline anisotropy
(MCA) as well as for calculations of the MCA constants. In particular,
the dependence  of the induced magnetic torque on different types of
atomic displacements is analyzed.    
\end{abstract}

\pacs{71.15.-m,71.55.Ak, 75.30.Ds}
\maketitle

\section {Introduction.}

While the ground state of magnetic materials is reasonably well described
within the spin density functional theory (SDFT) based first-principles approach, the Heisenberg model
is a tool giving access to the finite temperature and non-equilibrium
magnetic properties, making use of Monte Carlo \cite{PMS+10} or
spin-dynamics simulations \cite{SHNE08}, which are successfully applied
both to materials with robust local magnetic moment as well as to
metallic systems often treated as non-Heisenberg.  
In this content it is important to note that the exchange coupling
parameters $J_{ij}$ of the Heisenberg Hamiltonian are fully determined
by the crystal structure  
and electronic structure of a material, and can be estimated within the
framework of SDFT \cite{USK94,LKAG87,ALU+08,HBB08} for different systems,
although the leading mechanisms of the exchange interactions in these
materials may be different, depending on their structure and composition.  
A very efficient non-relativistic approach for calculations of the exchange
coupling parameters, based on the magnetic force
theorem (MFT), was suggested by Lichtenstein et al. \cite{LKAG87}, 
giving an explicit expression on the basis of the multiple scattering
formalism.  
Corresponding extensions of this computational scheme  
are now available to account for the full tensorial form 
of the interaction parameters \cite{USPW03,EM09a} as well as their 
extension to a multi-site  formulation  \cite{BSL19,MPE20}.

However, a description of magnetic properties based on a spin
Hamiltonian, in general, is incomplete, as it does not take into account
spin-lattice or magnetoelastic interactions. For some materials, a
corresponding contribution to the Hamiltonian may be neglected because
of a negligible spin-lattice coupling (SLC) 
while this is not the case for systems with strong spin-lattice
interactions which may be responsible for various interconnected
magnetic and structural properties. 
This concerns, for instance, a structural transformation accompanying
the magnetic ordering transition observed in non-collinear
antiferromagnets $R$MnO$_3$\cite{LPK+08} (where $R$ is a    
rare-earth element),  CuCrO$_2$ \cite{KON+09}, CuCrS$_2$
\cite{RBR+09}, and AgCrS$_2$ \cite{SMMS09}, as well as a formation of  
collinear order with complex structure in triangular lattice
antiferromagnets exhibiting strong geometrical frustration because of
antiferromagnetic nearest-neighbor exchange interactions
\cite{TMH+94,WV08}. 
A sufficiently strong spin-lattice coupling  may be responsible for
the magnon-phonon hybridization leading to mutual modifications of both
the magnon and phonon spectra, that have been found in experiments 
on the non-collinear antiferromagnets CuCrO$_2$ \cite{POL+16} and
(Y,Lu)MnO$_3$ \cite{OLN+16,KPLP19}. 
Recent investigations on ultra-fast demagnetization
\cite{BMDB96,VGL+12,GWY+16} demonstrate the importance of SLC for 
angular momentum transfer between the spin and lattice subsystems,
which may play a crucial role for the ultra-fast demagnetization
\cite{FTI+17,NK18,MKL19,DAS+19}.   
Some phenomena determined by magnon-phonon coupling are expected
to be useful for various applications, e.g. in spintronics.
This for example holds for the inverse Edelstein effect which implies spin to
charge current conversion with a spin current generated by a surface
acoustic wave in a ferromagnetic layer via magnon-phonon coupling
\cite{XPA+18}. Here one can also mention the possibility to drive
efficiently magnetic bubble domain walls, skyrmions  
and magnetic vortices by magnetoelastic waves \cite{OKB+15}, which is of
great practical 
importance for insulating materials when compared to metallic
systems where domain walls can be moved by spin currents.
A rapidly growing interest is also in the optical switching of the
magnetization driven by spin-lattice coupling \cite{JBH+17,SDS+21}.

Thus, the growing interest in various magnetic properties and
phenomena driven by SLC, motivates 
combined molecular-spin dynamics (MSD) simulations
\cite{MWD08,PEN+16,DAS+19,AN19,SER+21} 
that give at least access to the most central aspects 
of the above mentioned interesting and challenging experiments and
phenomena. 
So far, most of the investigations on the magneto-elastic properties of
materials have been performed on the basis of the phenomenological
continuous-field theory \cite{Kit58,Sch60,GM96} using  parameters
derived from experiment. 
 On the other hand, a practical scheme to calculate SLC parameters on the basis of 
  electronic structure calculations
   has been suggested by Hellsvik et al.\ \cite{HTM+19} and Sadhukhan et 
   al.\ \cite{SBK+22} for a system with an atom moved gradually from its
   equilibrium position.    
In the following we present an alternative scheme 
that treats changes to the spin configuration 
and atomic positions on the same level by
   applying a corresponding extension to the Lichtenstein formula \cite{MPL+22}.
This allows to derive closed expressions for the 
atomic SLC parameters in a coherent way up to any order, followed by
MSD simulation \cite{WLK+22} making use of these parameters.

Furthermore, we present a scheme to calculate the site-diagonal SLC
parameters characterizing local magnetic anisotropy induced by a lattice
distortion. 
It follows the approach suggested by Wang et al.\ \cite{WWWF96}, giving
access to the magnetic anisotropy via the calculation of the magnetic
torque, which accounts for all contributions to the magneto-crystalline
anisotropy (MCA).  Moreover, it allows to calculate in an efficient way all 
MCA constants entering the spin Hamiltonian. A more general expression was
worked out by Staunton et al. on the basis of multiple scattering theory
\cite{SSB+06}. Below we use a similar idea to calculate the MCA-like
contributions in the spin-lattice Hamiltonian.
It should be mentioned that in a complementary work \cite{LMP+23} we
consider in addition the role of the classical dipole-dipole interaction
for the SLC. Furthermore, this work presents and discusses further
numerical results for various 2D and 3D materials, in particular the magnetic
films and compounds which magnetic properties are strongly determined by
prominent magnetic frustrations and spin-lattice interactions.

\section {Intersite spin-lattice interactions}

\subsection{Spin-lattice Hamiltonian}

To describe  the coupling of the
spin and spatial degrees of freedom i.e.\
between the spin and lattice subsystems
we adopt an atomistic approach 
and start with the phenomenological spin-lattice Hamiltonian
%
\begin{eqnarray}
{\cal H}_{SLC} &=& 
                   -  \sum_{\substack{i,j,\alpha,\beta \\ k,\mu}} {\cal J}_{ij,k}^{\alpha
     \beta,\mu} 
                   e_i^{\alpha}e_j^{\beta}  
                   u^{\mu}_k
  -  \sum_{\substack{i,j \\ k,l}}  {\cal J}_{ij,kl}^{\alpha
     \beta,\mu\nu}  
                   e_i^{\alpha}e_j^{\beta}  
     u^{\mu}_k u^{\nu}_l  \nonumber \\
  && + \sum_{\substack{i,\alpha,\beta \\ k, \mu}}
                        {\cal K}^{\alpha\beta,\mu}_{i,k} e_i^{\alpha}
                        e_i^{\beta}u_k^\mu  +
  \sum_{\substack{i,\alpha,\beta \\ l,k,\mu\nu}}
                        {\cal K}^{\alpha\beta,\mu,\nu}_{i,kl} e_i^{\alpha}
     e_i^{\beta}u_k^\mu u_l^\nu  \;,  \nonumber \\                  
\label{eq:Hamilt_extended_magneto-elastic}
\end{eqnarray}
%
that can be seen as an
extension of the standard Heisenberg spin Hamiltonian.
Accordingly, the spin and lattice degrees of freedom
are represented by the orientation vectors $\vec e_{i(j)}$ of the magnetic
moments $\vec m_{i(j)}$ on the site $i(j)$, and atomic displacement vectors
$\vec u_{k(l)}$ for the atomic site $k(l)$.
In Eq.\ (\ref{eq:Hamilt_extended_magneto-elastic})
we omitted the spin-spin coupling (SSC) terms as well as the
elastic interaction term represented by the interatomic 
force constants \cite{HTM+19}, as we focus here on
the SLC parameters and their properties.
Moreover, the spin-lattice coupling has been restricted to three and four-site
terms in  Eq.\ (\ref{eq:Hamilt_extended_magneto-elastic}) (terms 1 and
2)  ${\cal J}_{ij,k}^{\alpha\beta,\mu}$ and ${\cal
  J}_{ij,kl}^{\alpha\beta,\mu\nu}$, described
in  tensorial form as relativistic effects are taken into account.
The parameters ${\cal K}^{\alpha\beta,\mu}_{i,k}$ and ${\cal
  K}^{\alpha\beta,\mu\nu}_{i,kl}$ characterize the local magnetic
anisotropy arising on site $i$ due to displacements of surrounding atoms
$k$ and $l$.
The Hamiltonian in Eq.\ (\ref{eq:Hamilt_extended_magneto-elastic})
that  is similar in form to that suggested by  Hellsvik et al. \cite{HTM+19}
obviously provides a suitable basis for advanced MSD simulations.

\subsection{Calculation of the $J^{\alpha\beta,\mu}_{ij,k}$ parameters}

In previous works expressions for  the non-relativistic and relativistic
exchange coupling parameters $J_{ij}$  \cite{LKAG87}
or $J_{ij}^{\alpha\beta}$ \cite{USPW03,EM09a}, respectively,   
have been derived by mapping the  free energy landscape ${F}(\{\vec e_i\})$
obtained from first-principles electronic structure calculations
on the Heisenberg spin Hamiltonian.
Here we follow the same strategy by mapping the
free energy landscape ${F}(\{\vec e_i\},\{\vec u_i\})$
by accounting for its dependency on the spin configuration $\{\vec e_i\}$
as well as  atomic displacements $\{\vec u_i\}$ on the same footing.
Making use of the magnetic force theorem
the change in free energy $ \Delta {F} $
induced by changes of the spin configuration  $\{\vec e_i\}$
with respect to a suitable reference system
and simultaneous finite  atomic displacements $\{\vec u_i\}$ 
may be written in terms of corresponding changes
to the single-particle energies:
%
\begin{eqnarray}
  \Delta {F} &=&   -\int^{E_F}dE \, \Delta N(E) \; ,
\label{eq:Free_energy}
\end{eqnarray}
where $E_F$ is the Fermi energy and $\Delta N(E)$ is the change to the
integrated  density of states (NOS)  $N(E)$.

As exploited before \cite{LKAG87,USPW03,EM09a},
$\Delta N(E)$
can be evaluated in a very efficient way via
the so-called Lloyd formula when  
 the underlying electronic structure is described 
by means of the multiple scattering or KKR (Korringa-Kohn-Rostoker)
formalism (see Appendix \ref{App:MST})\cite{EKM11}.
Adopting this approach one has:
%
\begin{eqnarray}
 \Delta {F} &=&  -\frac{1}{\pi} \mbox{Im\, Tr\,} \int^{E_F}dE\,
                      \left(\mbox{ln}\, \underline{\underline{\tau}}(E) - \mbox{ln}\, \underline{\underline{\tau}}^{0}(E)\right) \; , 
\label{eq:Free_energy-2}
\end{eqnarray}
with 
$ \underline{\underline{\tau}}^{(0)}(E)$ the so-called scattering path operator,
where the double underline indicates matrices with respect to  site and
spin-angular momentum indices \cite{EKM11}.
Within the KKR formalism these super matrices 
characterizing the 
reference ($\underline{\underline{\tau}}^{(0)}(E)$) and perturbed  ($\underline{\underline{\tau}}(E)$)
systems, respectively, are given by
%
\begin{eqnarray}
  \underline{\underline{\tau}}^{(0)}(E) &=&
          \left[\underline{\underline{m}}^{(0)}(E) - \underline{\underline{G}}(E)\right]^{-1}  \; ,
\label{eq:tau}
\end{eqnarray}
with   $\underline{\underline{G}}(E)$  the structure Green function and 
$\underline{\underline{m}}^{(0)}(E) =  [ \underline{\underline{t}}^{(0)}(E)  ]^{-1}$
the inverse of the corresponding
site-diagonal scattering matrix that carries
all site-specific information depending on
 $\{\vec e_i\}$ and  $\{\vec u_i\}$
\cite{EKM11}.
 
Considering a ferromagnetic reference state
($\hat{e}_i= \hat{e}_z$)  with all atoms
in their equilibrium positions ($\vec{u}_i=0$) the perturbed state is characterized
by finite  spin tiltings $\delta \hat{e}_i$
and finite  atomic displacements of the atoms $\vec{u}_i$ for the  sites $i$.
Writing for site $i$
the resulting changes in the inverse t-matrix
as   $\Delta^s_{\alpha} \underline{m}_i =
\underline{m}_i(\delta \hat{e}_i^\alpha) - \underline{m}^0_i $
and  $\Delta_{\nu}^u \underline{m}_i    =
\underline{m}_i({u}_i^\nu) - \underline{m}^0_i $
allows to replace the integrand in Eq.\ (\ref{eq:Free_energy-2}) by:
%
\begin{eqnarray}
  \mbox{ln} \,\underline{\underline{\tau}}
- \mbox{ln} \,\underline{\underline{\tau}}^0
  &=& - \mbox{ln}\left(1 +
       \underline{\underline{\tau}}\,[\Delta^s_{\alpha} \underline{{m}}_i + \Delta_{\nu}^u
                        \underline{{m}}_j +  ... ] \right) \; ,
\label{eq:tau2}
\end{eqnarray}
where all site-dependent changes in the spin configuration  $\{\vec e_i\}$
and atomic positions  $\{\vec u_i\}$
are accounted for in a one-to-one manner by the
various terms on the right hand side. This implies in particular that the
matrices $\Delta^{s}_{\alpha} \underline{{m}}_i $ and $\Delta^{u}_{\nu} \underline{{m}}_i $
in Eq.\ (\ref{eq:tau2}) are site-diagonal and have non-zero blocks only
for site $i$.
Assuming small tilting and displacement amplitudes leading in turn to
small changes of inversed scattering matrix $\Delta^s_{\alpha}
\underline{{m}}_i$ and $\Delta_{\nu}^u
\underline{{m}}_i$, a Taylor series expansion for the logarithm function in Eq.\
(\ref{eq:tau2})
gives access to the terms having different order with respect to the spin
tilting and atomic displacement. 
Making use of the magnetic force theorem, these blocks may be written 
in terms of the 
 spin tilting $\delta \hat{e}_i^\alpha$
and  atomic displacements of the atoms ${u}_i^\mu$ 
together with the corresponding 
 auxiliary matrices $\underline{T}^{\alpha}_i $
and ${\cal U}_{i}^{\mu}$, respectively,
as: 
\begin{eqnarray}
  \Delta^s_{\alpha} {\underline{m}}_i & = &  \delta \hat{e}^\alpha_i\,
       \underline{T}^{\alpha}_i \, \label{eq:linear_distor-s} \\
  \Delta^{u}_{\mu} \underline{m}_i & = & 
  u^\mu_i  \underline{\cal U}_{i}^{\mu} 
                                         \label{eq:linear_distor}
\end{eqnarray}
%
which represent the terms linear with respect to perturbations $\delta
\hat{e}_i^\alpha$ and ${u}_i^\mu$ (for more details see Appendix
\ref{App:UtU+}). 
Inserting these expressions into Eq.\ (\ref{eq:tau2})
and the result in turn into Eq.\ (\ref{eq:Free_energy-2})
allows in a straight forward way  to calculate the parameters
of the spin-lattice Hamiltonian  as the derivatives of the free energy
with respect to tilting angles and displacements. This way, accounting
for the 'minus' sign in the Hamiltonian in Eq.\
(\ref{eq:Hamilt_extended_magneto-elastic}), and using the third- and
fourth-order terms of the Taylor series in Eq.\ (\ref{eq:tau2}), one obtains 
for the SLC parameters up to fourth order the general expressions
%
\begin{eqnarray}
  {\cal J}^{\alpha\beta,\mu}_{ij,k} &=& - \frac{\partial^3 {\cal
                                F}}{\partial e^\alpha_i \,\partial e^\beta_j
                                        \, \partial u^{\mu}_k} = - \frac{1}{2\pi} \mbox{Im\, Tr\,}
                                  \int^{E_F}dE \,\, \nonumber \\
   && 
                          \times  \Bigg[   \underline{T}^{\alpha}_i \,\underline{\tau}_{ij}\,
                                  \underline{T}^{\beta}_j \,
                                 \, \underline{\tau}_{jk}\,
                                \underline{\cal U}^{\mu}_k \,\underline{\tau}_{ki} \nonumber \\
   & &                           +  \underline{T}^{\alpha}_i
                                   \,\underline{\tau}_{ik}\,
       \underline{\cal U}^{\mu}_k \,\underline{\tau}_{kj}
                                  \underline{T}^{\beta}_j \,
                                 \, \underline{\tau}_{ji}\Bigg]\,
       \label{eq:Parametes_linear}
\end{eqnarray}
and
\begin{eqnarray}
  {\cal J}^{\alpha\beta,\mu\nu}_{ij,kl} &=&  - \frac{\partial^4 {\cal
                                F}}{\partial e^\alpha_i \,\partial e^\beta_j
                                \, \partial u^{\mu}_k \partial
                                            u^{\nu}_l} = \frac{1}{2\pi} \mbox{Im\, Tr\,}
      \int^{E_F}dE \, \,  \nonumber \\
  & &  \times \Bigg[         \underline{T}^{\alpha}_i \,\underline{\tau}_{ij}\,
                                  \underline{T}^{\beta}_j \,
                                 \, \underline{\tau}_{jk}\,
      \underline{\cal U}^{\mu}_k \,\underline{\tau}_{kl}
      \,\underline{\cal U}^{\nu}_l\, \underline{\tau}_{li} \, \nonumber \\
    & &+      \underline{T}^{\alpha}_i \,\underline{\tau}_{ij}\,
                                  \underline{T}^{\beta}_j \,
                                 \, \underline{\tau}_{jl}\,
      \underline{\cal U}^{\mu}_l \,\underline{\tau}_{lk}
      \,\underline{\cal U}^{\nu}_k\, \underline{\tau}_{ki} \, \nonumber \\
    & &+      \underline{T}^{\alpha}_i \,\underline{\tau}_{il}\,
      \underline{\cal U}^{\nu}_l \,\underline{\tau}_{lj}
                                  \underline{T}^{\beta}_j \,
                                 \, \underline{\tau}_{jk}\,
      \underline{\cal U}^{\mu}_k \, \underline{\tau}_{ki} \, \nonumber \\
    & &+      \underline{T}^{\alpha}_i \,\underline{\tau}_{ik}\,
      \underline{\cal U}^{\mu}_k \,\underline{\tau}_{kj}
                                  \underline{T}^{\beta}_j \,
                                 \, \underline{\tau}_{jl}\,
      \underline{\cal U}^{\nu}_l \, \underline{\tau}_{li} \, \nonumber \\
   & &+   \underline{T}^{\alpha}_i \,\underline{\tau}_{ik}\,
      \underline{\cal U}^{\mu}_k \,\underline{\tau}_{kl} \, \underline{\cal U}^{\nu}_l\,
       \underline{\tau}_{lj} \,            \underline{T}^{\beta}_j \,
                                 \, \underline{\tau}_{ji}\,
      \, \nonumber \\
    & &+     \underline{T}^{\alpha}_i \,\underline{\tau}_{il}\,
      \underline{\cal U}^{\nu}_l \,\underline{\tau}_{lk} \, \underline{\cal U}^{\mu}_k\,
       \underline{\tau}_{kj} \,            \underline{T}^{\beta}_j \,
                                 \, \underline{\tau}_{ji}\,     \, \Bigg] \,,
\label{eq:Parametes_quadratic}
\end{eqnarray}
that supply the basis for corresponding calculations of
the SLC parameters \cite{MPL+22}. In the following these terms will be
called three- and four-site SLC terms, respectively, even if the  site
indices are identical. 
Note that the site-diagonal parameters, e.g. ${\cal
  J}^{\alpha\beta,\mu}_{ii,k}$ and  ${\cal J}^{\alpha\beta,\mu\nu}_{ij,ii}$,
may be contributed by the terms determined by single-site scattering matrix
corrections which are not only linear with respect to $\delta
\hat{e}_i^\alpha$ and ${u}_i^\mu$ (see Eqs.\ (\ref{eq:linear_distor-s})
and (\ref{eq:linear_distor})), but also quadratic, i.e., $\delta
\hat{e}^\alpha_i\,\delta \hat{e}^\beta_i\,
\underline{T}^{(2)\alpha\beta}_i$ (see for instance \cite{USPW03}) and 
$u^\mu_iu^\nu_i\, \underline{\cal U}_{i}^{(2)\mu\nu}$ (see Appendix
\ref{App:UtU+}), respectively. 

\begin{figure}[h]
 \includegraphics[width=0.4\textwidth,angle=0,clip]{SLC_Fe_Jij-jj_xx-uxux_lmax3.eps}\,(a)
 \includegraphics[width=0.4\textwidth,angle=0,clip]{SLC_Fe_Jij-jj_xy-uxux-2_lmax3.eps}\,(b)
\includegraphics[width=0.22\textwidth,angle=0,clip]{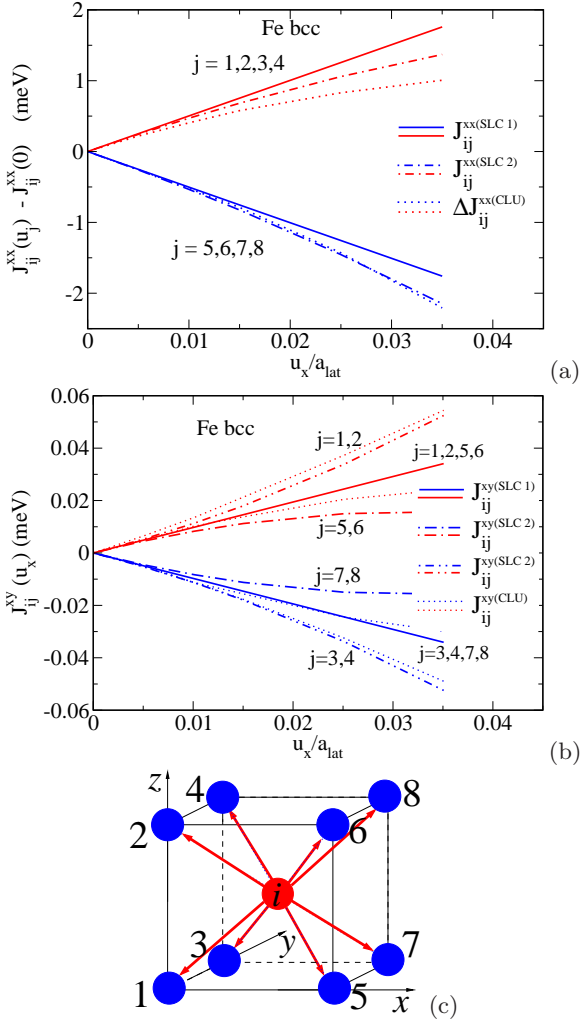}\,(c)
 \caption{\label{fig:quadratic}
The corrections of the diagonal (a) and
off-diagonal (b) exchange parameters, $\Delta J^{xx}_{ij}(u^x_i)$ and
$\Delta J^{xy}_{ij}(u^x_i)$ (dotted lines), respectively due to an atomic
displacement $u^x_i$ of atom $i$ along the $x$-axis, calculated for
bcc Fe. The results are compared with those based on the SLC parameters
multiplied by the corresponding displacement  ${J}^{xx (SLC 1)}_{ij} = {\cal J}^{xx,x}_{ij,i} \cdot
u^x_i$ and ${J}^{xy (SLC 1)}_{ij} = {\cal J}^{xy,x}_{ij,i} \cdot u^x_i$ (solid lines),
as well as ${J}^{xx (SLC 2)}_{ij} = {\cal J}^{xx,x}_{ij,i} \cdot
  u^x_{i} + {\cal J}^{xx,xx}_{ij,ii} \cdot u^x_{i}u^x_{i}$ (a)
  and ${J}^{xy (SLC 2)}_{ij} = {\cal J}^{xy,x}_{ij,i} \cdot
u^x_{i} + {\cal J}^{xy,xx}_{ij,ii} \cdot u^x_{i}u^x_{i}$ (b)
(dashed-dotted lines) plotted as a function of displacement amplitude. 
(c) Labeling of the nearest neighbor atoms for a bcc lattice:
  a displaced atom $i$ at the center is surrounded by atoms $j$
  with  $\vec{R}_{ij} = \vec{R}_{j} - \vec{R}_{i}$.   
  }     
 \end{figure}

In order to check the numerical results for three- and four-site SLC
parameters, $\underline{\cal J}_{ij,k}$, $\underline{\cal J}_{ij,kl}$,
obtained using the expressions in Eq.\ (\ref{eq:Parametes_linear}) and
(\ref{eq:Parametes_quadratic}), auxiliary calculations have been
performed delivering information about the changes 
of the exchange coupling parameters occurring due to a displacement of
one atom from its equilibrium position.
 For this purpose, the two-site SSC parameters
 $J^{\alpha\beta}_{ij}(u^\mu_{i})$ have been calculated for a cluster
 composed of 27 atoms, embedded into a bcc Fe lattice (see 
Ref.\ \onlinecite{EKM11}), with the
central atom $i$ displaced by $u^\mu_{i}$ along the $x$ direction, i.e. $\vec u_{k}||\hat x$.
Taking first- and second-order derivatives of $J^{\alpha\beta}_{ij}(u^\mu_{i})$ w.r.t.\
$u^\mu_{i}$ (assuming $\mu = x$) in the limit of $u^\mu_{i} = 0$ obviously allows a
direct comparison to $ {\cal J}^{\alpha\beta,\mu}_{ij,i} $ and $ {\cal
  J}^{\alpha\beta,\mu\mu}_{ij,ii} $. Alternatively, one may multiply  $ 
{\cal J}^{\alpha\beta,\mu}_{ij,i} $ with  $u^\mu_{i}$  and compare this
with $J^{\alpha\beta}_{ij}(u^\mu_{i})$  for varying $u^\mu_{i}$  (also assuming $\mu = x$). 
Note that all these calculations are performed for a ferromagnetic (FM)
reference system with its magnetization $\vec M$ along the $z$-axis,
i.e.\ $\vec M||\hat z$.
The corresponding diagonal $\alpha = x$, $\beta=x$ and off-diagonal, $\alpha = x$,
$\beta=y$ tensor elements in spin subspace, seen as a 
function of the displacement $u^x_{i}$, are compared in  
Fig.\ \ref{fig:quadratic}, (a) and (b), respectively.
 For the diagonal terms shown in Fig.\ \ref{fig:quadratic}(a) one can
 see two groups of data belonging to atoms $i=$ 1 -- 4 and $i=$ 5 -- 8,
 respectively, (see Fig.\  \ref{fig:quadratic} (c)) that have opposite sign.
 Obviously, a rather good agreement between the data for
 $J^{{xx}}_{ij}(u^x_{i})$ and ${\cal J}^{{xx},x}_{ij,i} \cdot
 u^x_{i}$ is found for a small amplitude of the displacement.
The same applies also for the off-diagonal terms shown 
in Fig.\ \ref{fig:quadratic} (b).
%
When the displacement amplitude increases, the diagonal elements
$J^{{xx}}_{ij}(u^x_{j})$ deviate from the linear dependence on the
displacement, increasingly with its amplitude.
On the other hand, the off-diagonal terms show an additional
splitting up and down away from a linear variation, both for the curves showing
positive and negative sign. This can be attributed to the impact of higher-order
terms with respect to the displacement.
To check this, additional calculations have been performed accounting
for second-order contributions to the exchange coupling tensor,
quadratic with respect to the displacements giving access to the
combination ${\cal J}^{{\alpha\beta},x}_{ij,i} \cdot 
u^x_{i} + {\cal J}^{{\alpha\beta},xx}_{ij,ii} \cdot u^x_{i}u^x_{i}$. 
The second term is calculated using the expression for the SLC
tensor elements given by Eq.\ \ref{eq:Parametes_quadratic},
assuming $k=i$ and $l=i$. 
Note however, that in this case (i.e. $k=i$ and $l=i$) an 
additional second-order contribution $ {\cal
  J}^{(2)\alpha\beta,\mu\nu}_{ij,ii}$, has to be taken into account, 
represented by the expression
\begin{eqnarray}
  {\cal J}^{(2)\alpha\beta,\mu\nu}_{ij,ii} &=&  -\frac{1}{2\pi} \mbox{Im\, Tr\,}
                                  \int^{E_F}dE \,\, \nonumber \\
   && 
                          \times  \Bigg[   \underline{T}^{\alpha}_i \,\underline{\tau}_{ij}\,
                                  \underline{T}^{\beta}_j \,
                                 \, \underline{\tau}_{ji}\,
                                \underline{\cal U}^{(2)\mu\nu}_i \,\underline{\tau}_{ii} \nonumber \\
   & &                           +  \underline{T}^{\alpha}_i
                                   \,\underline{\tau}_{ii}\,
       \underline{\cal U}^{(2)\mu\nu}_i \,\underline{\tau}_{ij}
                                  \underline{T}^{\beta}_j \,
                                 \, \underline{\tau}_{ji}\Bigg]\,,
       \label{eq:Parametes_quadratic-3}
\end{eqnarray}
where $\underline{\cal U}^{(2)\mu}_i$ stems from the second order derivative
of the distorted matrix $\underline{m}_k$ with respect to the
displacement (see Appendix \ref{App:UtU+}), which includes the following two contributions
\begin{eqnarray}
  \Delta^{2,u}_{\mu\nu} \underline{m}_i & = & 
  u^\mu_i u^\nu_i (\underline{\cal U}_{i}^{(2a)\mu\nu} + \underline{\cal U}_{i}^{(2b)\mu\nu})
                                         \label{eq:linear_distor}
\end{eqnarray}
%
with 
\begin{eqnarray}
\underline{\cal U}_{i}^{(2a)\mu\nu} &=& -\bigg( \underline{U}^{\mu}(\hat{u}_i)
\underline{m}_i \underline{U}^{\nu}(\hat{u}_i) + \underline{U}^{\mu}(\hat{u}_i)
\underline{m}_i \underline{U}^{\nu}(\hat{u}_i)\bigg)\,, \nonumber \\  
 \underline{\cal U}_{i}^{(2b)\mu\nu}  & = & \bigg(
                                            \underline{\bar{U}}^{(2)\mu\nu}(\hat{u}_i)\underline{m}_i 
            + \underline{m}_i \,\underline{\bar{U}}^{(2)\mu\nu}(\hat{u}_i)\bigg)  
\label{eq:linear_distor}
\end{eqnarray}

The dependencies of the terms ${\cal J}^{xx,x}_{ij,i} \cdot
  u^x_{i} + {\cal J}^{xx,xx}_{ij,ii} \cdot u^x_{i}u^x_{i}$ 
  and ${\cal J}^{xy,x}_{ij,i} \cdot
  u^x_{i} + {\cal J}^{xy,xx}_{ij,ii} \cdot u^x_{i}u^x_{i}$ on the
  displacement are shown in
  Fig.\ \ref{fig:quadratic} (a) and (b), respectively, by dashed-dotted
  line, demonstrating their good agreement with $\Delta J^{xx}_{ij}(u^x_i)$
  and $\Delta J^{xy}_{ij}(u^x_i)$, respectively, calculated for an embedded
  cluster with a displaced atom in the center. 
  In addition, the dependence of the $\Delta
  J^{xy}_{ij}(u^x_i)$ parameter on the position of atom $j$ is determined
  by corresponding dependencies of the three-site and four-site parameters
  ${\cal J}^{xy,x}_{ij,i}$ and ${\cal J}^{xy,xx}_{ij,ii}$ presented in
  Table \ref{tab:table-xy_ij}.
\begin{table}[h!]
  \begin{center}
    \caption{
      The nearest-neighbor three-site  ${\cal 
        J}^{xy,x}_{ij,i}$  (a) and four-site ${\cal
        J}^{xy,xx}_{ij,ii}$ (b) SLC parameters (meV/a.u. and
      meV/(a.u.)$^2$ units, respectively). The results are given for
      the shifted atom $i$ at the center and the nearest neighbor sites
      $j$ (see  Fig.\ \ref{fig:quadratic} (c)).}  
    \label{tab:table-xy_ij}
    \begin{tabular}{l|c|c|c|c|c|c|c|c}
      \hline
   $i$ & $1$ & $2$  & $3$ & $4$
                        & $5$  & $6$ & $7$ & $8$ \\
      \hline
  a  &  0.182  &  0.182 & -0.182 & -0.182 & 0.182 & 0.182 & -0.182 & -0.182   \\
\hline
  b    & 0.506  & 0.506 &-0.506 & -0.506 &-0.506 &-0.506 & 0.506  &  0.506  \\
\hline
    \end{tabular}
  \end{center}
\end{table}

The different sign of the parameters ${\cal J}^{xy,x}_{ij,i}$ (line (a) in
the table) indicates a different slope for the two linear branches  ${\cal
  J}^{xy,x}_{ij,i} \cdot  u^x_{i}$, that can be observed for two groups of atoms $j$ in
Fig. \ref{fig:quadratic} (b), positive for $j = \{1,2,5,6\}$ and 
negative for $j = \{3,4,7,8\}$. The parameters ${\cal
  J}^{xy,xx}_{ij,ii}$ given in line (b) of  Table \ref{tab:table-xy_ij}
characterize the curvature of the function ${\cal J}^{xy,xx}_{ij,ii}
\cdot u^x_{i}u^x_{i}$ quadratic with respect to the displacement.
Within the first group of neighbors, the curvature is positive for $j =
1,2$ and negative for $j =5,6$, while within the second group it is positive for $j =
7,8$ and negative for $j =2,4$. As a consequence, the quadratic contribution
results in a splitting of the linear branches in line with the results for ${\cal
  J}^{xy}_{ij}(u^{x}_i)$ obtained from selfconsistent
calculations for embedded clusters with a displaced atom $i$.

The anti-symmetric part of the off-diagonal three-site SLC tensor
elements 
can be seen as the Dzyaloshinskii-Moriya interaction $D^z_{ij}$ induced
by the symmetry-breaking displacement 
of atom $k$, i.e., ${\cal D}^{z,\mu}_{ij,k} =  \frac{1}{2}({\cal
  J}^{xy,\mu}_{ij,k} - {\cal J}^{yx,\mu}_{ij,k})$, that occurs despite the
conventional DMI represented by $D^\alpha_{ij}$ vanishes for the non-distorted bcc Fe
lattice due to inversion symmetry. 
The same applies also for the four-site SLC parameters ${\cal D}^{z,\mu\nu}_{ij,kl}$.

\begin{figure}[t]
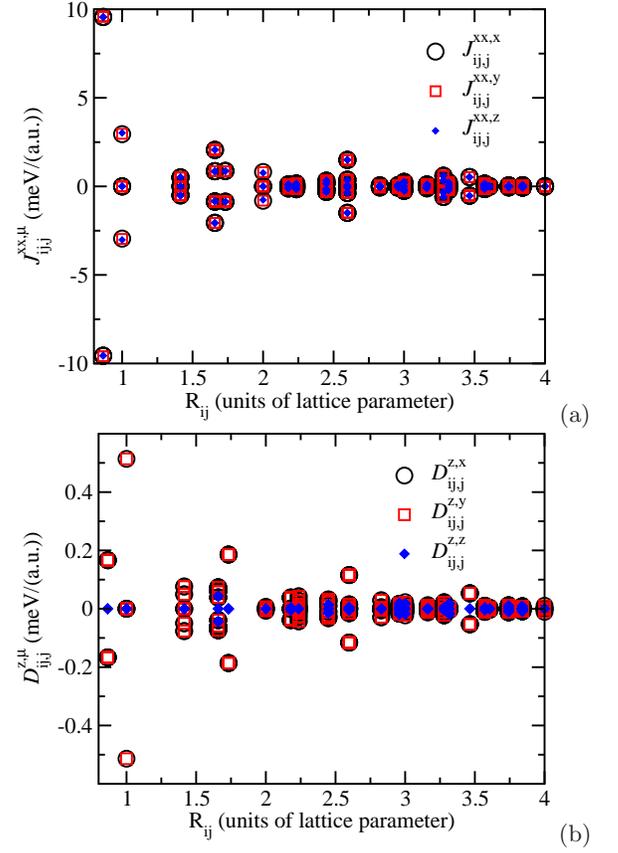

\includegraphics[width=0.4\textwidth,angle=0,clip]{Fe_Jij-uj_vs_Rij.eps}\,(a)
\includegraphics[width=0.4\textwidth,angle=0,clip]{Fe_Dij-uj_vs_Rij.eps}\,(b)
\caption{\label{fig:Fe-Jijij_diag-u_cmp} The SLC parameters for bcc Fe (a) diagonal
  ${\cal J}^{xx,\mu}_{ij,j}$, and (b) DMI-like ${\cal D}^{z,\mu}_{ij,j}$,
  represented as a function of the interatomic distance $R_{ij}$. 
}     
\end{figure}


For illustration, the three-site and four-site SLC parameters
characterizing the spin-spin coupling between the non-displaced atom $i$ and
displaced atom $j$ have been calculated for FM ordered bcc Fe.
Fig.\ \ref{fig:Fe-Jijij_diag-u_cmp} represents the diagonal ${\cal   J}^{xx,\mu}_{ij,j}$ and DMI-like
 ${\cal D}^{z,\mu}_{ij,j}$ SLC parameters plotted as a  
function of the interatomic distance ${R}_{ij}$, for different
directions $u_x$, $u_y$, $u_z$ of the displacement.
Both figures (a) and (b) look symmetric with respect to a sign inversion
of the SLC parameters as a consequence of the above mentioned spitting
of the parameters into two groups with opposite sign. Of course,
these groups behave differently depending on the direction of the displacement
$\vec{u}_j$. Moreover, for the DMI-like SLC parameters, one can
see different magnitudes of the ${\cal D}^{z,z}_{ij,j}$ parameters
(i.e. for $\vec{u}_j$ parallel to the DMI vector) when compared to ${\cal
  D}^{z,x}_{ij,j}$ and ${\cal D}^{z,y}_{ij,j}$ with the displacements
perpendicular to the DMI vector.

The SLC tensor with the elements ${\cal
  J}^{\alpha\alpha,\mu}_{ij,k}$,  multiplied
by spin tiltings $\delta \hat{e}_{i(j)}$ on sites
$i$ and $j$ represents a force acting on the atom on site $k$. These
forces can lead to a structure instability induced by magnetic order in
a system as mentioned in the introduction.

\begin{figure}[h]
\includegraphics[width=0.18\textwidth,angle=0,clip]{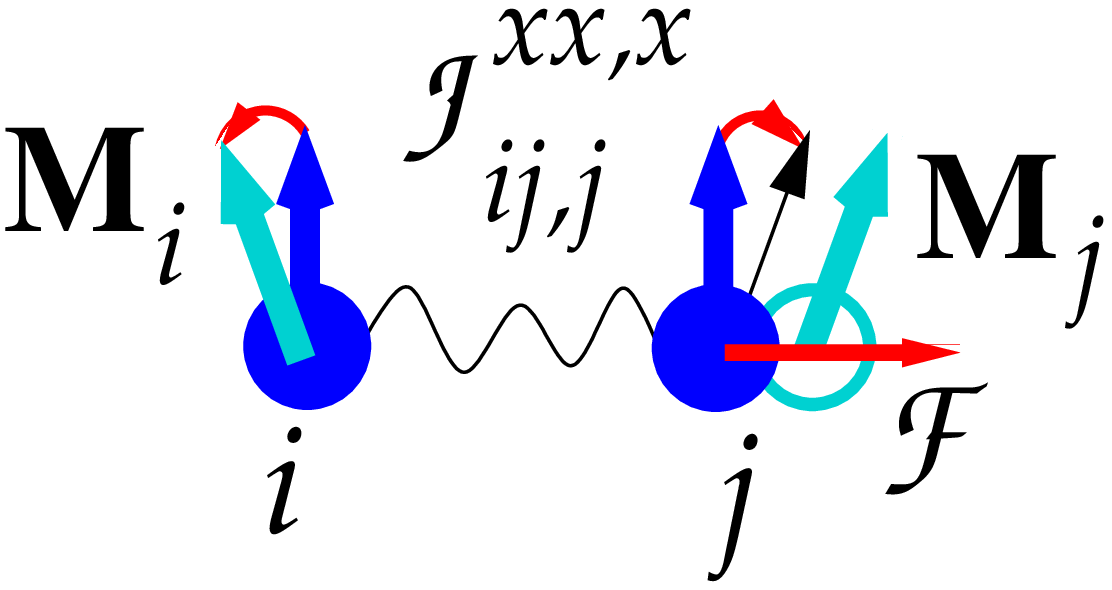}\,\,\,
\includegraphics[width=0.26\textwidth,angle=0,clip]{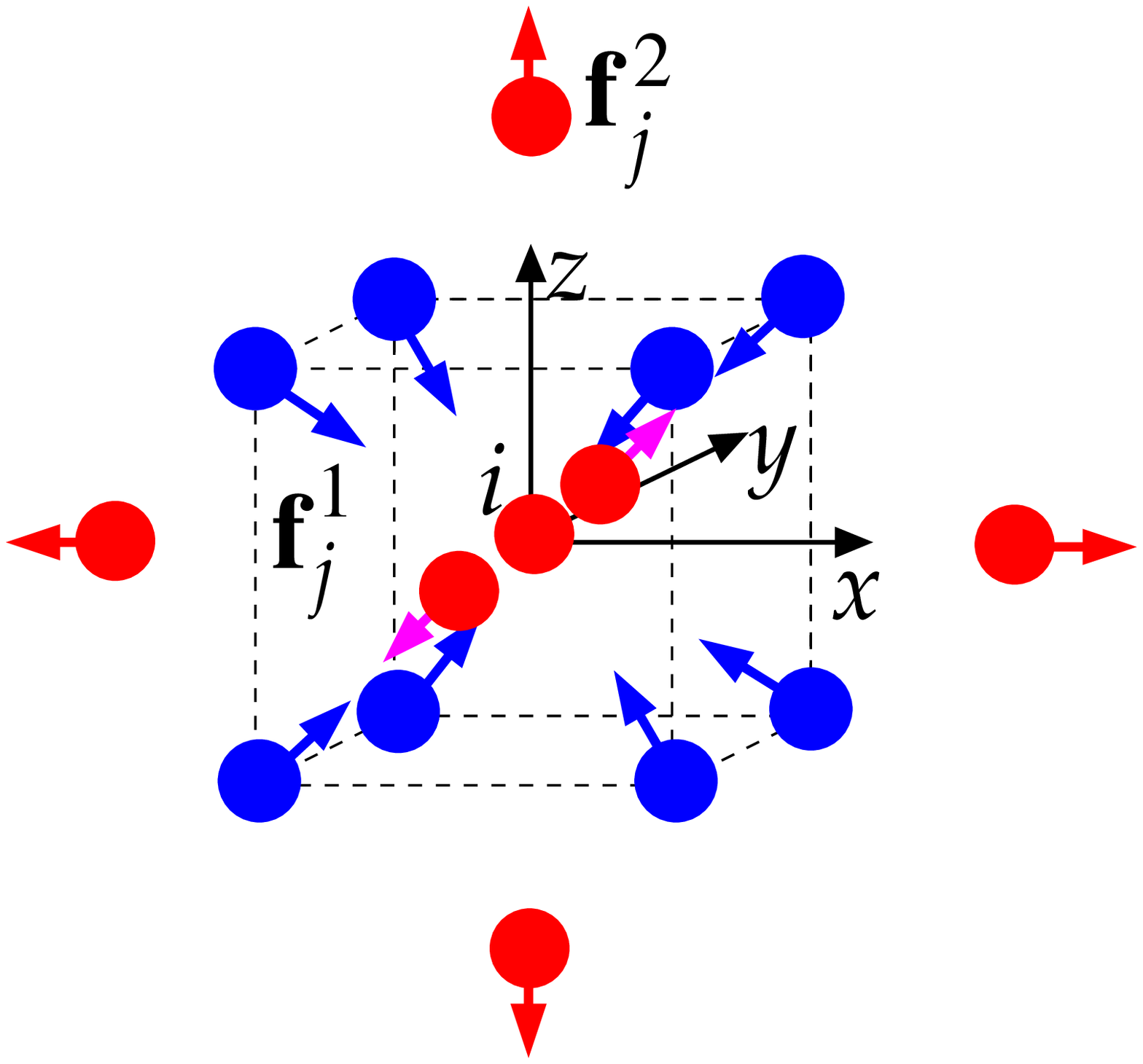}\,(a)
\includegraphics[width=0.15\textwidth,angle=0,clip]{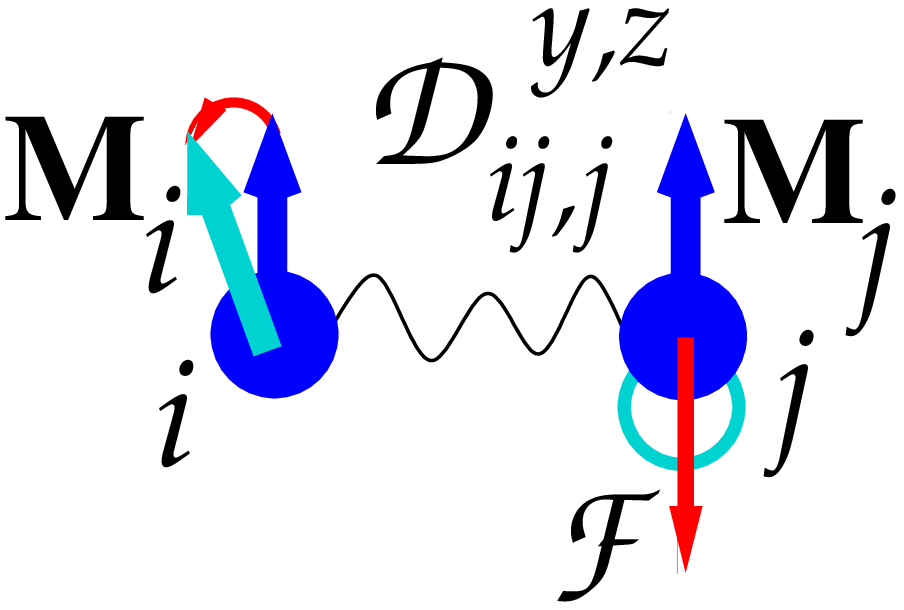}\,\,\,
\includegraphics[width=0.22\textwidth,angle=0,clip]{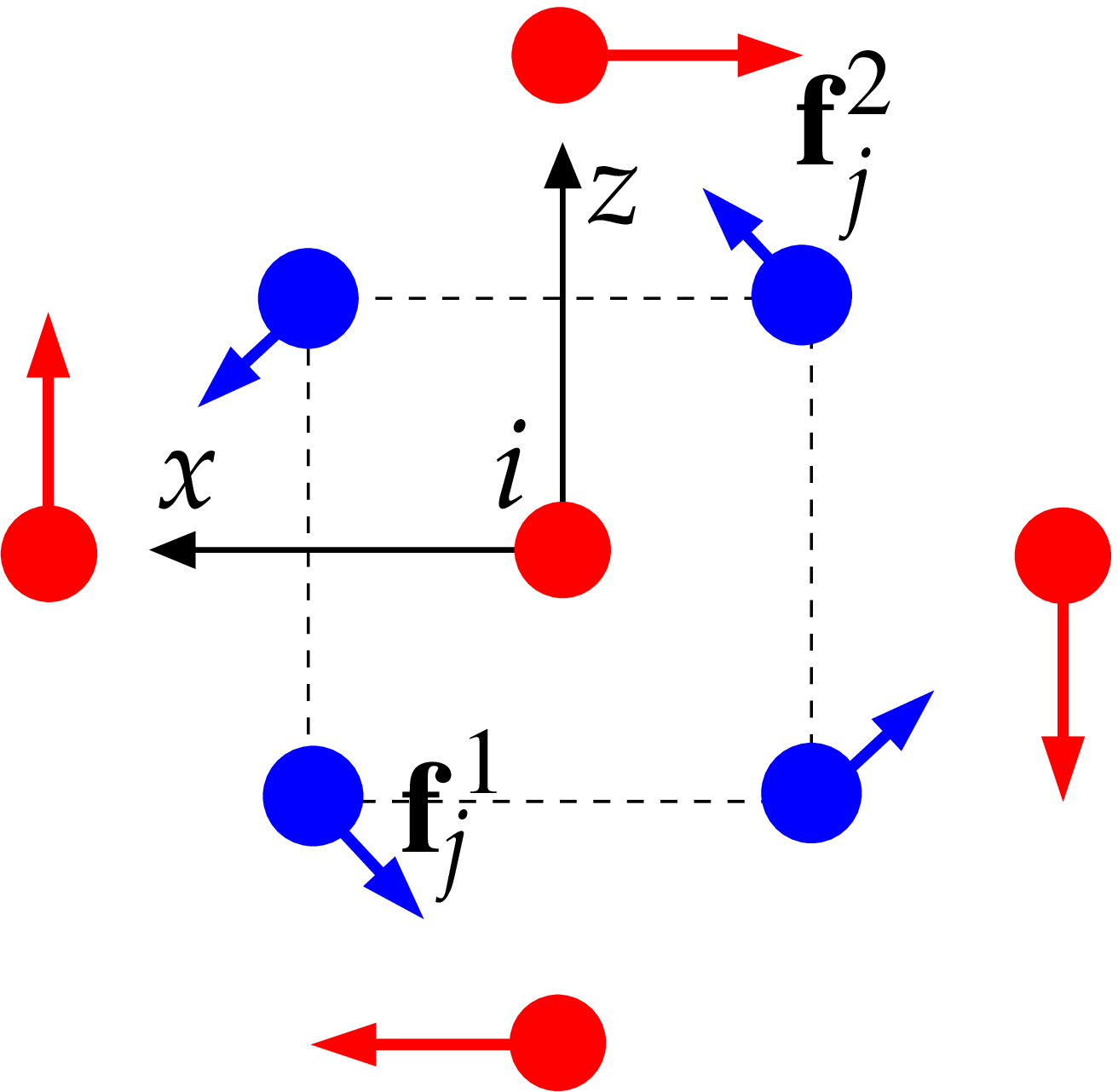}\,(b)
\caption{\label{fig:Fe-Jij-Dij_force}
  The quantities  $\vec{f}_j = -\sum_\mu{\cal
    J}^{{\rm dia-s},\mu}_{ij,j}\hat{n}_\mu$ (a) and  $\vec{f}_j = -\sum_\mu {\cal
    D}^{y,\mu}_{ij,j} \hat{n}_\mu$ (b) ($\hat{n}_x = \hat{x}$,
  $\hat{n}_y = \hat{y}$, $\hat{n}_z = \hat{z}$) associated with the
  symmetric diagonal and DMI-like SLC, respectively, characterizing the
  forces on atoms $j$ induced by spin tilting on site $i$ via the spin-lattice 
  coupling in bcc Fe with the magnetization along $z$ axis.
The left panel shows schematically the SLC mediated force $\vec{\cal F}$ on atom
$j$ (red arrow) induced by tilting of the spin moments on sites $i$ and $j$ (shown
in light blue color), and
vice versa, the spin tiltings induced due to the displacements of the atom
on site $j$ (light blue circle). In the right panel, the 
  arrows show the directions of the forces for the first and second
  atomic shells, blue and red, respectively. In the case (a) the forces
  are directed along the lines connecting the interacting
  atoms, with $\vec{f}_j = 9.56(\pm 1,\pm 1,\pm 1)$ for the first shell
  and  $\vec{f}_j = -2.9(0,0,\pm 1)$, $\vec{f}_j =
  -2.9(0,\pm 1,0)$, $\vec{f}_j = -2.9(\pm 1,0,0)$ for the second
  shell. In the case (b) the forces are perpendicular to the lines
  connecting the interacting atoms, with $\vec{f}^1_j = 0.16(
  \pm 1,0,\pm 1)$ for the first shell and $\vec{f}^2_j = 0.5(0,0,\pm 1)$,
  $\vec{f}^2_j = 0.5(\pm 1,0,0)$ for the second shell. }     
\end{figure}

Let us discuss the forces $\vec{\cal F}$ generated due to the symmetric
diagonal ${\cal J}^{{\rm dia-s},\mu}_{ij,j} = \frac{1}{2}({\cal J}^{xx,\mu}_{ij,j}
+ {\cal J}^{yy,\mu}_{ij,j})$ and DMI-like ${\cal
  D}^{\alpha,\mu}_{ij,j}$ spin-lattice interactions, as shown in the
left pannels of Fig.\ \ref{fig:Fe-Jij-Dij_force}(a) and (b), respectively.  
Considering FM bcc Fe with the magnetization direction along the $z$
axis, the corresponding forces generated due to a rotation of the spin  
moments on sites $i$ and $j$, $\hat{e}_{i(j)} \approx \hat{z} + \delta
\hat{e}^x_{i(j)}$ have the components $-{\cal J}^{{\rm dia-s},\mu}_{ij,j}\delta
e_i^{x} \delta e_j^{x} $ and $-{\cal  D}^{y,\mu}_{ij,j} (\hat{e}_i \times
\hat{e}_j)_y = -{\cal  D}^{y,\mu}_{ij,j}(\hat{e}^z_i \delta
\hat{e}^x_{j} - \hat{e}^z_j \delta \hat{e}^x_{j})$, respectively.
In Fig.\ \ref{fig:Fe-Jij-Dij_force} the arrows show the quantities
$\vec{f}_j = -\sum_\mu{\cal  J}^{{\rm dia-s},\mu}_{ij,j}\hat{n}_\mu$ (a)
and $\vec{f}_j -\sum_\mu{\cal D}^{y,\mu}_{ij,j}\hat{n}_\mu$ (b) 
(with the unit vectors $\hat{n}_x = \hat{x}$, $\hat{n}_y = \hat{y}$,
$\hat{n}_z = \hat{z}$), which determine corresponding forces $\vec{F}_j$ on
the atoms at site $j$, arising due to spin tiltings within the $x-z$ plane. 
The results are presented for two atomic shells around atom $i$.
As one can see, the forces originated from the diagonal symmetric
elements of the SLC tensor are directed along the lines connecting the
interacting atoms. This may lead to a lattice distortion being the result of
a competition with the  elastic forces between the atoms.
On the other hand, as one can see in Fig.\  \ref{fig:Fe-Jij-Dij_force} (b), the forces
originating from the antisymmetric elements of the SLC tensor, i.e. the
 DMI-like SLC parameters, are perpendicular to the lines connecting the 
 interacting atoms, creating a mechanical torque on the lattice dependent
 on the magnetic configurations, that can lead to an angular momentum
 transfer upon the magnon-phonon scattering events \cite{MPL+22,GC15,
   RSBD20}. 

\begin{figure}[h]
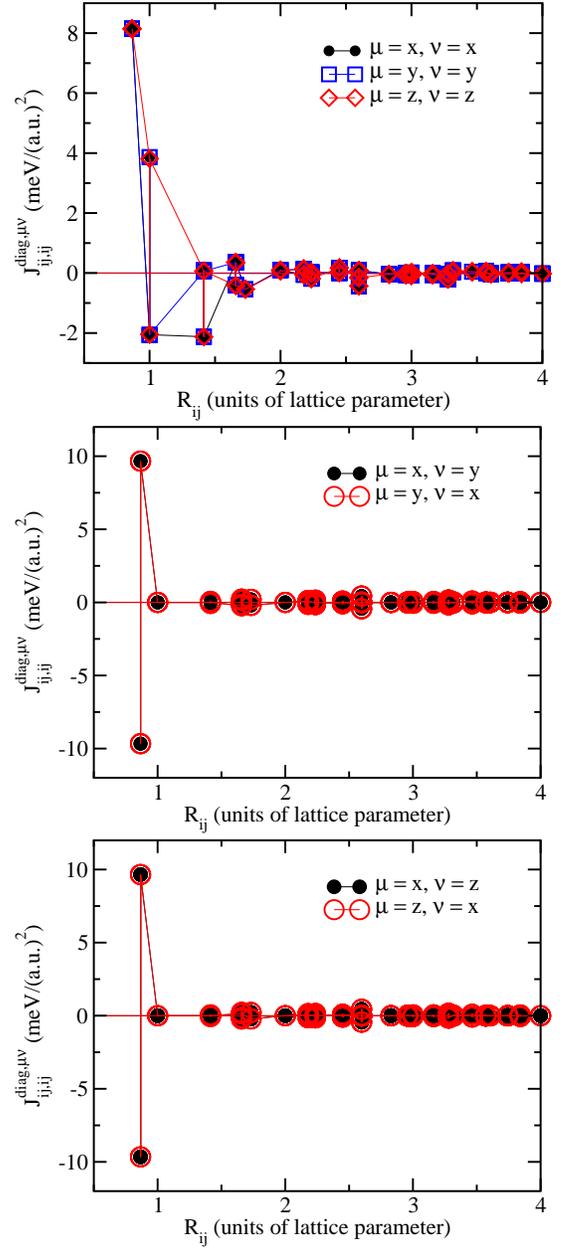

\includegraphics[width=0.4\textwidth,angle=0,clip]{Fe_CMP_SLC_diag_ui-uj-diag.eps}
\includegraphics[width=0.4\textwidth,angle=0,clip]{Fe_CMP_SLC_diag_ui-uj-off-diag-xy.eps}
\includegraphics[width=0.4\textwidth,angle=0,clip]{Fe_CMP_SLC_diag_ui-uj-off-diag-xz.eps}
\caption{\label{fig:Fe-Jijij_diag-uu_cmp} The SLC parameters ${\cal J}^{{\rm
      dia-s},\mu\nu}_{ij,ij}$, diagonal
  with respect to spin $\alpha = \beta$ and (a) diagonal with respect to
  displacement indices $\mu = \nu$ (${\cal J}^{{\rm
      dia-s},\mu\mu}_{ij,ij}$),  and off-diagonal with respect to
  displacement indices, $\mu \neq \nu$, (b) for $\mu =
  \{x,y\}$ and $\nu = \{x,y\}$ and (c) for $\mu = \{x,z\}$ and $\nu =
  \{x,z\}$, represented as a function of interatomic distance $R_{ij}$. 
}     
\end{figure}
\begin{figure}[h]
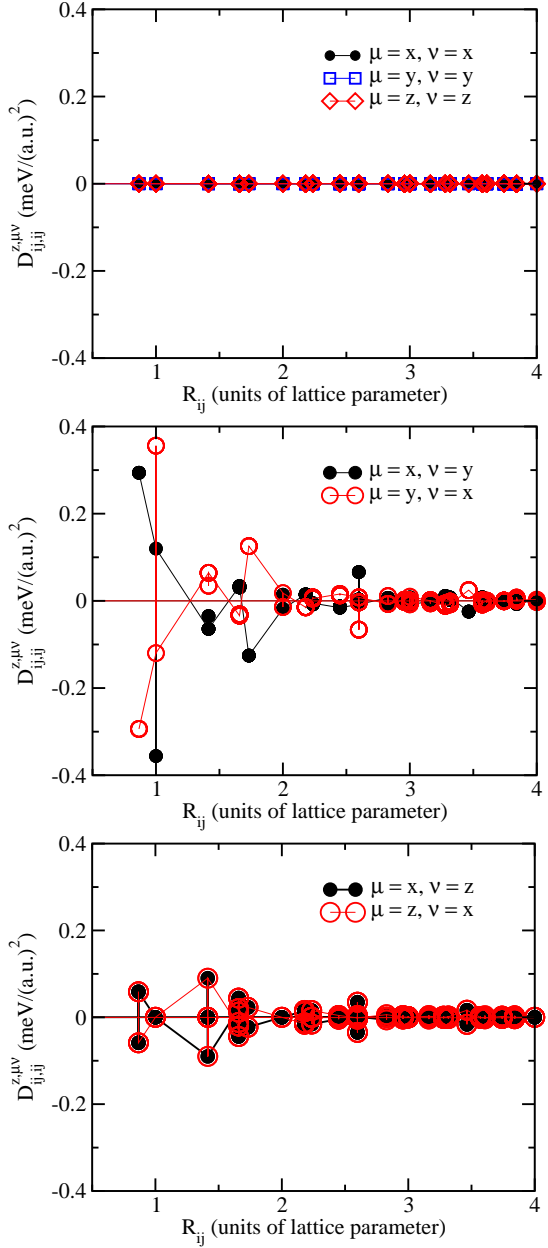

\includegraphics[width=0.4\textwidth,angle=0,clip]{Fe_CMP_SLC_DMI-z_ui-uj-diag.eps}
\includegraphics[width=0.4\textwidth,angle=0,clip]{Fe_CMP_SLC_DMI-z_ui-uj-off-diag-xy.eps}
\includegraphics[width=0.4\textwidth,angle=0,clip]{Fe_CMP_SLC_DMI-z_ui-uj-off-diag-xz.eps}
\caption{\label{fig:Fe-Dijij-uu_cmp} The DMI-like SLC parameters ${\cal
    D}^{z,\mu\nu}_{ij,ij}$ diagonal with respect to displacement indices
  $\mu = \nu$ (a), and off-diagonal with respect to
  displacement indices, $\mu \neq
  \nu$, (b) for $\mu = \{x,y\}$ and $\nu = \{x,y\}$ and (c) for $\mu =
  \{x,z\}$ and $\nu = \{x,z\}$, 
  represented as a function of interatomic distance $R_{ij}$.
}     
\end{figure}

Figs.\ \ref{fig:Fe-Jijij_diag-uu_cmp} and \ref{fig:Fe-Dijij-uu_cmp}
represent the four-site SLC parameters, ${\cal J}^{{\rm dia-s},\mu\nu}_{ij,ij}
= \frac{1}{2}\sum_k({\cal J}^{xx,\mu\nu}_{ij,ij} + {\cal J}^{yy,\mu\nu}_{ij,ij})$
and  ${\cal D}^{z,\mu\nu}_{ij,ij}$, respectively, plotted as a function
of the interatomic distance $R_{ij}$. In this case both atoms $i$ and $j$
are assumed to be displaced from the equilibrium. One can see a
dominating nearest-neighbor coupling  ${\cal  J}^{{\rm dia-s},\mu\nu}_{ij,ij}$
in the case of  $\mu \neq \nu$, while the ${\cal
  J}^{{\rm dia-s},\mu\mu}_{ij,ij}$  coupling has a comparable strength for
several neighboring shells.
The SOC-driven DMI-like 
parameters are about two orders of magnitude smaller, and are strongly
determined by the local symmetry depending on the directions of the displacements
and the direction of the $\vec{\cal D}^{\mu\nu}_{ij,ij}$ vector. In particular,
 as is shown in Fig.\ \ref{fig:Fe-Dijij-uu_cmp} (a), the  ${\cal
   D}^{z,\mu\nu}_{ij,ij}$ component is equal to zero for the
 displacement of atoms $i$ and $j$ along the same direction.
 
\medskip


\section{Site-diagonal spin-lattice coupling parameters}

\subsection{Phenomenology: MCA-like spin-lattice Hamiltonian}

In addition to the interatomic spin-lattice interaction, the SLC
Hamiltonian in Eq.\ (\ref{eq:Hamilt_extended_magneto-elastic}) includes
 also a contribution to be seen as a counterpart to the magnetic
 anisotropy in the spin Hamiltonian
\begin{eqnarray}
  {\cal H}_{MA} &=& \sum_{i,\alpha,\beta} K^{\alpha\beta}_{i} e_i^{\alpha} e_i^{\beta}
  + \sum_{i,\alpha,\beta,\gamma,\delta} K^{\alpha\beta\gamma\delta}_{i} e_i^{\alpha}
                    e_i^{\beta}e_i^{\gamma}e_i^{\delta} +  ... \;
                    \nonumber \\
\label{eq:MCA-Hamilt_1}
\end{eqnarray}
with the non-vanishing terms determined by the symmetry of the crystal.
The corresponding MCA-like terms in the spin-lattice
Hamiltonian in Eq.\ (\ref{eq:Hamilt_extended_magneto-elastic})
characterize contributions to the magnetic anisotropy at 
any site, arising due to a displacement of surrounding atoms, breaking
the local symmetry of the crystal.

The induced magnetic anisotropy energy in the
Hamiltonian is characterized by the anisotropy constants,
which may include different
contributions discussed in the literature, controlled by 
dipole-dipole interactions and the spin-orbit coupling (SOC) \cite{Lee55}. 
When comparing the dipole-dipole contribution to the anisotropy and
magnetostriction observed experimentally, Lee \cite{Lee55} points out 
that it represents only a small part of the observed values.
This led him to the conclusion that the magnetoelastic constants
are primarily determined by SOC.

Therefore we focus here on the SOC-driven spin-lattice coupling
responsible for the local magnetic anisotropy induced by a lattice
distortion breaking the local symmetry in the system. Dealing with the
atomistic spin-lattice Hamiltonian keeping the lowest-order terms
linear with respect to the atomic displacements according to the
expression
\begin{eqnarray}
  {\cal H}_{\rm{me-MA}} &=& \sum_{i,k}\sum_{\mu} ({\cal K}_{i,k}^{xx,\mu} u^{\mu}_k e_i^x e_i^x
                     +{\cal K}_{i,k}^{yy,\mu} u^{\mu}_k e_i^y e_i^y \nonumber \\
              && +  {\cal K}_{i,k}^{zz,\mu} u^{\mu}_k  
                     e_i^z e_i^z  +   {\cal K}_{i,k}^{xy,\mu} u^{\mu}_k e_i^x
                 e_i^y \nonumber \\ 
              && + {\cal K}_{i,k}^{xz,\mu} u^{\mu}_k e_i^x e_i^z +
                {\cal K}_{i,k}^{yz,\mu} u^{\mu}_k e_i^y e_i^z   )   \;, 
\label{eq:Totque_2}
\end{eqnarray}
we will discuss below an approach providing the basis for calculations
of the SLC parameters of the Hamiltonian 
in Eq.\ (\ref{eq:Hamilt_extended_magneto-elastic}) on a first-principles level.
Some contributions to the expression in Eq.\ (\ref{eq:Totque_2}) have been
discussed already \cite{MPL+22}, 
which correspond to the site-diagonal SLC tensor $\underline{\cal
  J}_{ii,k}$, both, diagonal, e.g. ${\cal J}^{{\rm dia-a},\mu}_{ii,k}
=  \frac{1}{2}({\cal J}^{\alpha\alpha,\mu}_{ii,k} - {\cal
  J}^{\beta\beta,\mu}_{ii,k})$ and off-diagonal, e.g.
${\cal   J}^{{\rm off-s},\mu}_{ii,k} =  \frac{1}{2}({\cal
  J}^{xy,\mu}_{ii,k} + {\cal J}^{yx,\mu}_{ii,k})$, terms.
 However, there are further contributions which have to be
    taken into account, similar to those discussed in Ref.\ \onlinecite{USPW03}
considering various contributions to the MCA.
In particular, one should mention the so-called non-local contribution
associated with the anisotropy of the three-site SLC 
${\cal J}^{{\rm dia-a},\mu}_{ij,k}
=  \frac{1}{2}({\cal J}^{\alpha\alpha,\mu}_{ij,k} - {\cal
  J}^{\beta\beta,\mu}_{ij,k})$, similar to the so-called non-local
contribution $\frac{1}{2}({\cal J}^{\alpha\alpha}_{ij} -
{\cal J}^{\beta\beta}_{ij})$ to the uniaxial magnetic anisotropy
discussed in  Ref.\ \cite{USPW03}.



As an alternative, we are going to use a scheme based on the magnetic 
torque \cite{SSB+06} (see Appendix \ref{App:Torque}), to get access 
to the parameters of the MA-SLC  
Hamiltonian in Eq.\ (\ref{eq:Totque_2}).
Focusing on the terms ${\cal K}^{\alpha z,\mu}_{i,k}  e_i^\alpha e_i^z
u^{\mu}_k$ , the SLC parameters are directly connected to the
effective field determined as
\begin{eqnarray}
 \vec{H}_{i,\rm{eff-SLC \alpha z}} &=& -\frac{\partial }{\partial e_i^{\alpha}} {\cal
  H}_{\rm{me-MA}} |_{\theta=0}  \;, 
\label{eq:Torque_magneto-elastic_4}
\end{eqnarray}
and can be calculated as follows
\begin{eqnarray}
  {\cal K}^{\alpha z,\mu}_{i,k} e_i^{z} &=& \frac{\partial}{\partial^2 e_i^{\alpha}\partial u_k^{\mu}} {\cal
                                            H}_{\rm{me-MA}} |_{\theta=0}
  \nonumber \\
  &=& -\frac{\partial}{\partial u_k^{\mu}}
\vec{H}_{i,\rm{eff-SLC \alpha z}}   \;.
\label{eq:Torque_magneto-elastic_3}
\end{eqnarray}

For the FM reference state with the equilibrium magnetization direction
along the $z$ axis, one has  
$e^x \approx \theta$, $e^y \approx \theta$ and $e^z \approx 1 -
\frac{1}{2}\theta^2 \approx 1$, assuming a small spin tilting $\theta$ from the equilibrium.
 This allows to redefine for the sake of convenience the SLC
  parameters as follows ${\cal K}^{\alpha
  z,\mu}_{i,k} e_i^z \to \tilde{\cal K}^{\alpha z,\mu}_{i,k}$ \cite{KA53,Kit58,GM96},
keeping in mind that the newly defined parameters are
antisymmetric with respect to time reversal and their original form
should be used in the dynamical equations.
The corresponding SLC parameters can be calculated via the first derivative
with respect to the spin direction, i.e. 
\begin{eqnarray}
\tilde{\cal K}^{\alpha z,\mu}_{i,k} &=& \frac{\partial}{\partial e_i^{\alpha}\partial u_k^{\mu}} {\cal
  H}_{\rm{me-MA}} |_{\theta=0} =  \frac{\partial}{\partial e_i^{\alpha} \partial u_k^{\mu}}
               {F}|_{\theta=0} \;.
\label{eq:Torque_magneto-elastic_3}
\end{eqnarray}

\subsection{Torque: First-principles approach}

As a starting point we use the ferromagnetic (FM) state as a reference
state and neglect for the moment all temperature effects, 
i.e.\ assume $T = 0$ K.
Instead of using the Lloyd formula, we represent the change
of free energy $\Delta {F}$ in terms of the Green function
$G_0(E)$ for the FM reference state, which is modified due to the perturbation.
Denoting the corresponding change in the Green function $\Delta
    G(E)$ and neglecting temperature effects one can write the change of
    the total energy:
\begin{eqnarray}
\Delta {F} & \approx &  -\frac{1}{\pi} 
\mbox{Im}\, \mbox{Tr}  \int^{E_F} dE \,
(E - E_F) \, \Delta G(E) \; ,
\label{Eq_DeltaF_1}
\end{eqnarray}
where $E_F$ is the Fermi energy.
Assuming that the perturbations are small, the induced 
change of the Green function can be represented 
by the following perturbative expansion
\begin{eqnarray}
  \Delta G(E) & = & G_0(E) \Delta V_m G_0(E) \nonumber\\
  && +  G_0(E) \Delta V_m G_0(E) \Delta V_m G_0(E) + ... \;, \nonumber\\
  && +  G_0(E) \Delta V_m G_0(E) \Delta V_u G_0(E) + ... \;,
\label{Eq_GF_expansion}
\end{eqnarray}
 where $\Delta V_m$  describes a perturbation due to the spin-tilting,
 and  $\Delta V_u$ is a perturbation due to a 
 lattice distortion in the system.
 We will deal with the first and third terms in  Eq.\
 (\ref{Eq_GF_expansion}). 
 
Substituting Eq.\ (\ref{Eq_GF_expansion}) into Eq.\
(\ref{Eq_DeltaF_1}) and using the sum rule $\frac{dG}{dE} = - GG$ for the
Green function, one obtains an expression for the total energy change
associated with the perturbations:
\begin{eqnarray}
\Delta {F} &=&   \frac{1}{\pi} \mbox{Im}\,\mbox{Tr} \int^{E_F} dE\, (E - E_F) \,
 \Delta V_m \, \frac{dG_0(E)}{dE} \nonumber \\ 
&&+ \frac{1}{\pi} \mbox{Im}\,\mbox{Tr} \int^{E_F} dE\, (E - E_F)  \nonumber \\ 
                &&
                   \times \, \Delta V_m
   G_0(E)  \Delta V_u \frac{dG_0(E)}{dE}  \;. 
\label{Eq_Free_Energy1}
\end{eqnarray}

We keep here only the first- and second-order terms that give access to the
magnetic torque for the crystal, without and with a lattice distortion, respectively.

By performing an integration by parts for the second equation in Eq.\ (\ref{Eq_Free_Energy1})
and taking into account that $ (E - E_F) \, \Delta
V_m \, G_0(E)|_{E = E_F} = 0$ and $ (E - E_F) \, \Delta
V_m \, G_0(E) \Delta V_u \, G_0(E)|_{E = E_F} = 0$, the free energy change
$\Delta {F}$ is given by:
\begin{eqnarray}
\Delta {F} &=&   -\frac{1}{\pi} \mbox{Im}\,\mbox{Tr} \int^{E_F} dE\, \Delta V \, G_0(E) \nonumber \\ 
&&-\frac{1}{\pi} \mbox{Im}\,\mbox{Tr} \int^{E_F} dE\, \Delta V_m
   G_0(E)  \Delta V_u G_0(E) \\
   &=& { F}^{(1)} + { F}^{(2)}   \;.
\label{Eq_Free_Energy2}
\end{eqnarray}
%

Representing the Green function in terms of the multiple scattering
formalism \cite{EKM11}, Eq.\ (\ref{Eq_Free_Energy2}) leads to the expression
\begin{eqnarray}
\Delta {F} &=&   -\frac{1}{\pi} \mbox{Im}\,\mbox{Tr} \int^{E_F}
                    dE\, {\langle\Delta \underline{V}_m \rangle} \, \underline{\tau}(E) \nonumber \\  
&&-\frac{1}{\pi} \mbox{Im}\,\mbox{Tr} \int^{E_F} dE\, {\langle\Delta \underline{V}_m\rangle}
   \underline{\tau}(E) {\langle\Delta \underline{V}_u\rangle} \underline{\tau}(E) \;.
\label{Eq_Free_Energy3}
\end{eqnarray}

Using for the matrix elements of perturbation
$\langle \Delta \underline{V}_m \rangle$ and $\langle \Delta \underline{V}_u \rangle$
the expressions (see Appendix \ref{App:UtU+})
\begin{eqnarray}
  {\langle \Delta \underline{V}_m \rangle} & = &  \sum_\alpha \delta \hat{e}^\alpha_i\,
                                           \underline{T}^{\alpha}_i \, \\
  {\langle \Delta \underline{V}_u \rangle} & = & \sum_\nu u^\nu_i  \underline{\cal U}_{i}^{\nu}                                         
\end{eqnarray}
and taking the derivatives $\frac{\partial
  {F}^{(1)}}{\partial e^\alpha}$ and  $\frac{\partial^2 {F}^{(2)}}{\partial
  e^\alpha \partial u^\mu}$, one obtains for the magnetic anisotropy constants
\begin{eqnarray}
  \tilde{K}^{\alpha z}_{i} &=&  \frac{\partial \Delta {F}}{\partial
  e_i^\alpha}\bigg|_{u=0} = - \frac{1}{\pi} \mbox{Im}\,\mbox{Tr} \int^{E_F}
                    dE\,  \underline{T}^{\nu}_i \, \underline{\tau}_{ii}(E)    \nonumber \\ 
\label{Eq_Free_Energy_deriv_K1}
\end{eqnarray}
and their counterparts in the spin-lattice Hamiltonian, 
\begin{eqnarray}
 \tilde{\cal K}^{\alpha z,\mu}_{i,k} &=&  \frac{\partial^2 \Delta {F}}{\partial
                   e_i^\alpha \partial u_k^\mu}\bigg|_{u=0}  \nonumber \\
 &=& - \frac{1}{\pi} \mbox{Im}\,\mbox{Tr}
                               \int^{E_F} dE\,  \underline{T}^{\mu}_i
   \underline{\tau}_{ik}(E) \underline{\cal U}_{k}^{\nu} \underline{\tau}_{ki}(E) \;.
\label{Eq_Free_Energy_deriv_K2}
\end{eqnarray}

\begin{figure}[b]
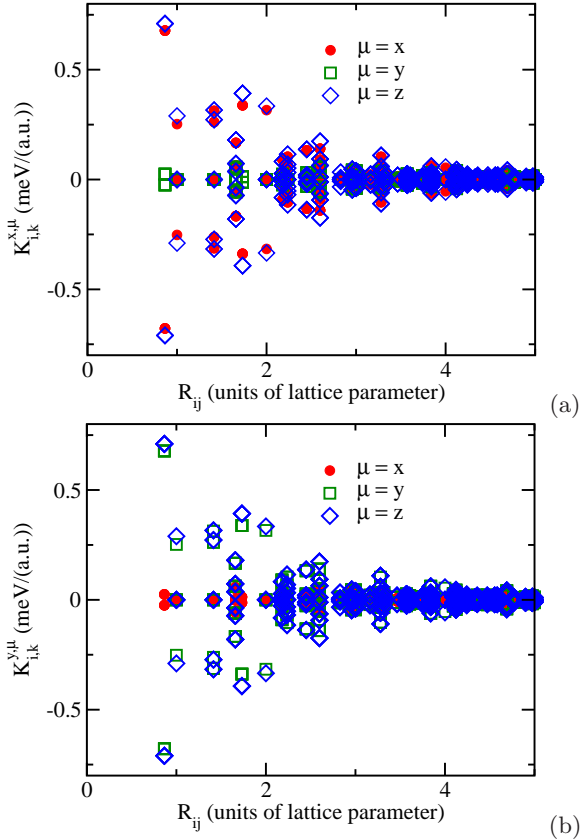

\includegraphics[width=0.4\textwidth,angle=0,clip]{Fe_CMP_SLC_torque-xx-x.eps}\,(a)
\includegraphics[width=0.4\textwidth,angle=0,clip]{Fe_CMP_SLC_torque-xy-x.eps}\,(b)
\caption{\label{fig:torque_ik_xx-x} The parameters  ${\cal K}_{i,k}^{x,\mu}$
  (a) and ${\cal K}_{i,k}^{y,\mu}$ (b)  calculated for bcc Fe,
  as a function of the distance $R_{ik}$. 
 }     
\end{figure}

The parameters ${\cal K}^{\alpha z,\mu}_{i,k}$ 
give access either to the
torque on a magnetic moment $ \vec{\cal T}_i = \hat{e}_i \times
\vec{H}_{eff}$ caused by the effective field induced by the 
displacements  $u_k^\mu$ of the atoms on sites $k$, i.e.
\begin{eqnarray}  
 \vec{H}_{eff,i}^\alpha &=& - \sum_{k,\mu} 
                   {\cal K}^{\alpha z,\mu}_{i,k} \hat{e}^z_i u_k^\mu \;,
\label{Eq_Torque_spin}
\end{eqnarray}
and characterizing the rate of change of spin angular momentum,
or to the mechanical torque, e.g. $\vec{\mathfrak{T}}^{ph}_k = \vec{u}_k \times
\vec{\cal F}_k$ created by the forces induced by spin tiltings
$\hat{e}^\alpha_i$  on sites $i$, i.e.,
\begin{eqnarray}  
 {\cal F}^\mu_k &=& - \sum_{i, \alpha}
                   {\cal K}^{\alpha z,\mu}_{i,k} \hat{e}^\alpha_i \hat{e}^z_i \;,
\label{Eq_Torque_lattice}
\end{eqnarray}
and contributing to the rate of change of the spin angular momentum of
phonons \cite{RSBD20}.
As an example, the parameters ${\cal K}^{\alpha z,\mu}_{i,k}$ and ${\cal
  K}_{i,k}^{yz,\mu}$ calculated for bcc Fe (with the magnetization
direction along $z$ axis) are plotted in
Fig. \ref{fig:torque_ik_xx-x} as a function of the distance $R_{ik}$,
for the three  displacements $\mu = x,y,z$. As one can see,
their absolute values are much smaller in the case of a displacement
perpendicular to the plane of magnetization rotation (i.e. for the $u_y$
component for tilting within the $x-z$ plane and the $u_x$ component for
tilting within the $y-z$ planes) when compared to the displacements 
within the plane.

As it was already pointed out, the displacement of any atom $k$ in the
system, obviously, breaks the local symmetry at a neighboring site $i$,
creating a corresponding contribution to the magnetic anisotropy and in turn
to a corresponding effective field and the torque on the magnetic moment on site $i$.
This torque depends on the local symmetry around the displaced atoms, as well as
on the direction of the magnetization with respect to the crystal lattice.
Using the phenomenological SLC Hamiltonian, one can see that the contributions
to the effective field caused by the displacement $u^\mu_k$,
which are associated with different SLC terms in the Hamiltonian,
have a different dependence on the magnetization direction.

\begin{figure}[t]
\includegraphics[width=0.45\textwidth,angle=0,clip]{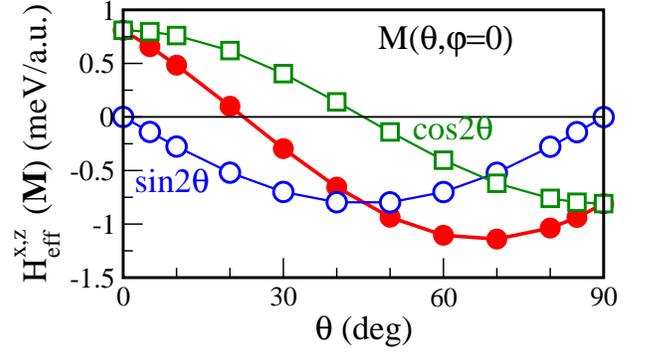}\,
\caption{\label{fig:torque_single} The dependence of the effective field (closed circles)
 in bcc Fe on the $\theta$ angle characterising the direction of the
  magnetization w.r.t. $\hat{z}$, in the presence of a single atom
  displaced along the  $\hat{z}$ direction.
  It is contributed  by the terms $\sim ({\cal K}_{i,k}^{{{x}}{{x}},z}
  - {\cal K}_{i,k}^{{{z}}{{z}},z})\mbox{sin}2\theta $ (open circles)
  and $\sim ({\cal K}_{i,k}^{{{x}}{{z}},z} + {\cal
    K}_{i,k}^{{{z}}{{x}},z})\mbox{cos}2\theta$ (open squares).
 }     
\end{figure}

To demonstrate this dependence, we consider FM ordered bcc Fe and focus
on the term $\sim {\cal K}_{i,k}^{xz,\mu}$. In the case of the magnetization
oriented along the crystallographic direction [001] and atoms $k$
displaced along the $\hat{z} || (0,0,1)$ direction, i.e. $\vec{u} =
u^z_k\hat{z}$, the corresponding effective magnetic field induced on
site $i$ is equal to ${\cal K}_{i,k}^{xz,z} e^z u^z_k$. It determines
the induced torque on the magnetic moment on site $i$
responsible for the formation of a noncollinear magnetic
structure driven by the lattice distortion. At the same time, the
induced effective field due to the terms $\sim {\cal K}_{i,k}^{xx,z} e^x_i 
u^z_k$ and ${\cal K}_{i,k}^{yy,z} e^y u^z_k$  for such a
geometry is equal to zero as
$e^x \sim \theta = 0$ and $e^y \sim \theta = 0$.
Rotating the frame of reference together with the magnetization (within the
$x-z$ plane by the angle $\theta$), keeping
$\hat{\tilde{z}} || \hat{m}$, the effective field is calculated via
the transformation $\underline{R}_{-\theta} \underline{A}
\underline{R}_{-\theta}^{-1}$ with $\underline{A}$ seen as the matrix
with the elements ${A}^{\alpha\beta} \sim {\cal K}_{i,k}^{\alpha\beta,z} u^z_k$. 
As a result, the non-vanishing effective field is given 
 by the expression  $H^{\tilde{x}}_i(u^z_k) = - \frac{\partial E}{\partial e^{\tilde{x}}_i} = 
-({\cal K}_{i,k}^{{\tilde{x}}{\tilde{z}},z} + {\cal K}_{i,k}^{{\tilde{z}}{\tilde{x}},z}){e}^{\tilde{z}}_i u^z_k$,
where
\begin{eqnarray}
 H^{\tilde{x}}_i(u^z_k)  &=& -\frac{1}{2}\bigg[({\cal
                                                K}_{i,k}^{{{x}}{{x}},z}
                                                - {\cal
                                                K}_{i,k}^{{{z}}{{z}},z})\mbox{sin}2\theta
 \nonumber \\
                                            &&    + ({\cal
                                                K}_{i,k}^{{{x}}{{z}},z}
                                                + {\cal
                                                K}_{i,k}^{{{z}}{{x}},z})\mbox{cos}2\theta
                                                \bigg] \;. 
\label{Eq_Anisotr_const_transformation}
\end{eqnarray}
%
A similar expression can also be found
for ${\cal K}_{i,k}^{{\tilde{z}}{\tilde{x}},z}$.
Fig. \ref{fig:torque_single} represents a particular example of the
effective field  $H^{\tilde{x}}_i(u^z_k)$ on site $i$ ($R_i = (0,0,0)$) in bcc Fe, which is
created due to a displacement of atom $k$, $R_k = a(0.5,0.5,0.5)$,
along the crystallographic direction $[001]$, assuming $|u^z_k| = 1$.
This field can now be calculated on a first-principles level,  using the
expression in Eq.\ (\ref{Eq_Free_Energy_deriv_K2}). The total field is shown
by the red solid line, which is a result of two contributions $\sim
\mbox{sin}2\theta$  and $\mbox{cos}2\theta$, shown by blue dashed and
green dashed-dotted lines, respectively.

\begin{figure}[t]
\includegraphics[width=0.45\textwidth,angle=0,clip]{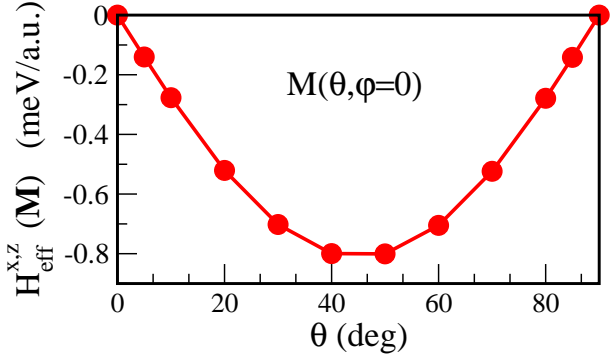}\,
\caption{\label{fig:torque_tetra} The dependence of the effective
   field (closed circles)
  in bcc Fe on the $\theta$ angle characterizing direction of the
  magnetization w.r.t. $\hat{z}$, in the presence of nearest-neighbor 
  displacements along  $\hat{z}$ direction, corresponding to a tetragonal
  distortion of the crystal.
%
 }     
\end{figure}

In the case of a tetragonal distortion, i.e. a displacement by $u_z$ of the atoms
at $a(\pm0.5,\pm0.5,0.5)$ and a displacement by $-u_z$ of the atoms at
$a(\pm0.5,\pm0.5,-0.5)$, the only non-zero contribution due to the induced
effective field is associated with the term $\frac{1}{2}({\cal
                                                K}_{i,k}^{{{x}}{{x}},z}
                                                - {\cal
                                                K}_{i,k}^{{{z}}{{z}},z})$,
i.e. $\sim \mbox{sin}2\theta$, that is shown in Fig.\
\ref{fig:torque_tetra}. Note that in this case the displacement amplitude
has been normalized by the factor $1/8$ to represent the energy (or
field) per one displaced atom. 

Note that the effective field considered here is coused by the SOC-induced
 anisotropic part of the exchange tensor, seen as a non-local
contribution to the magnetic anisotropy, that concerns also the
anisotropy induced by a lattice distortion (via SLC). One should point
out that these contributions stem from the DMI-like SLC given by the expression 
\begin{eqnarray}
 {H}^x_i (u^\mu_k) = \sum_{j,k,\mu} ( \hat{e}_j \times \vec{D}^{\mu}_{ij,k})_x u^\mu_k
  \label{fig:torque_DMI}
\end{eqnarray}
as well as the anisotropy of the diagonal elements of the SLC tensor
 $-\frac{1}{2}\sum_k({\cal J}^{xx,z}_{ij,k} - {\cal J}^{zz,z}_{ij,k})$.
The former one has a dependence on the magnetization direction
similar to that of the site-diagonal contribution related to the ${\cal 
  K}_{i,k}^{{{x}}{{z}},\mu} u^\mu_k$ MCA-like term, which is associated to
the DMI-like SLC $\vec{\cal D}^{z}_{ij,k}$. As one can see in Fig.\
\ref{fig:torque_tetra}, this contribution vanishes in the case of a
tetragonal distortion of the lattice as this deformation does not
break inversion symmetry.
The effective field associated with the diagonal anisotropy of
the SLC tensor is responsible for a uniaxial magnetic anisotropy.
It does not vanish in the case of the displacements shown in
Fig. \ref{fig:torque_tetra}, as well as in the case of a
tetragonal distortion, and is responsible for the non-local contribution
to the magnetic anisotropy discussed in the literature \cite{USPW03,MPB+11}.



\begin{figure}[b]
\includegraphics[width=0.4\textwidth,angle=0,clip]{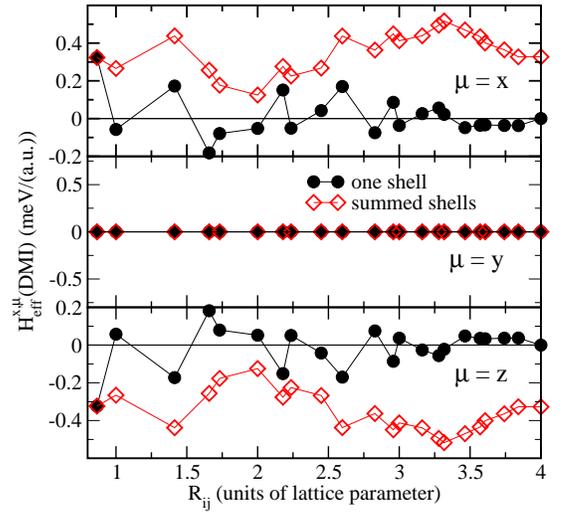}\,
\caption{\label{fig:TORQUE_Dij-uj_vs_Rij} The shell-resolved 
  effective field for bcc Fe contributed by the DMI-like parameters ${\cal
  D}^{y,\mu}_{ij,k}$, i.e. $- \sum_{j:|R_{ij}| \in d_n} {\cal
  D}^{y,\mu}_{ij,k}$  (closed symbols) up to $|\vec{R}_{ij}^{max}| =
  4a$, accounting for displaced nearest neighboring atom $k$. Open
  symbols represent the effective field as a function of $|\vec{R}_{ij}|$
  summed up over all contributions up to $|\vec{R}_{ij}|$, i.e.
  $ - \sum_{j: |R_{ij}| \leq d_n} {\cal D}^{y,\mu}_{ij,k}$. Top panel
  corresponds to $\mu = x$, middle panel to $\mu = y$, and bottom - to $\mu = z$.
 }     
\end{figure}

\begin{figure}[h]
\includegraphics[width=0.4\textwidth,angle=0,clip]{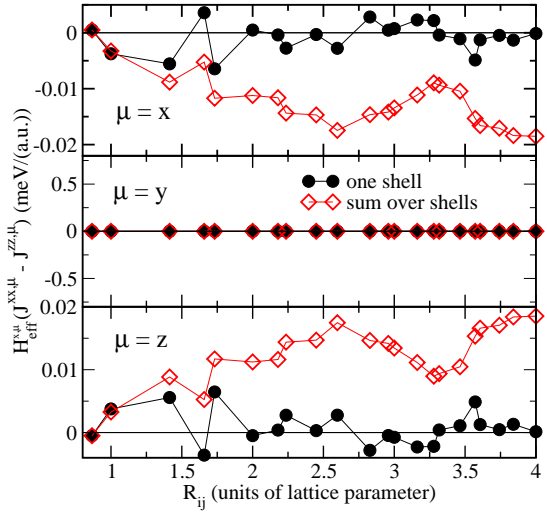}\,
\caption{\label{fig:TORQUE_diag-Jij-uj_vs_Rij} The shell-resolved 
  effective field for bcc Fe contributed by the diagonal anisotropic
  part of the SLC tensor $1/2({\cal J}^{xx,\mu}_{ij,k}
  - {\cal J}^{zz,\mu}_{ij,k})$  i.e. $-\sum_{j:|R_{ij}| \in d_n} 1/2({\cal J}^{xx,\mu}_{ij,k}
  - {\cal J}^{zz,\mu}_{ij,k})$  (closed symbols) up to $|\vec{R}_{ij}^{max}| =
  4a$, accounting for displaced nearest neighboring atom $k$. Open
  symbols represent the effective field as a function of $|R_{ij}|$
  summed up over all contributions up to $|\vec{R}_{ij}|$, i.e.
  $-\sum_{j: |R_{ij}| \leq d_n} 1/2({\cal J}^{xx,\mu}_{ij,k}
  - {\cal J}^{zz,\mu}_{ij,k}) $. The top panel
  corresponds to $\mu = x$, middle panel to $\mu = y$, and bottom - to $\mu = z$.
 }     
\end{figure}

Discussing the properties of the effective field determined by the three-site SLC parameters
(the same concerns also other multisite SLC parameters), one has to take
into account the ${\cal J}^{\alpha\beta,\mu}_{ij,k}$ parameters with $k
\neq i$ and $k \neq j$. Thus, their contribution to the effective field
(torque) at site $i$ was investigated for the case of a displaced nearest
neighboring atom $k$, but accounting for all SLC contributions
including $j \neq k$. 
 Fig. \ref{fig:TORQUE_Dij-uj_vs_Rij} shows shell-resolved
  DMI-SLC contributions to the effective field  ${H}^x_i (u^\mu_k)$  
induced by a displacement $u^\mu_{k}$ of the atom at $\vec{R}_{01} =
a(0.5,0.5,0.5)$. It is determined by a coupling via the term ${\cal
  D}^{y,\mu}_{ij,k}$ (according to Eq.\ (\ref{fig:torque_DMI})) 
of the central atom $i$ with all atoms $j$ within the
shell $n$ with the radius $d_{n}$ going up to $d_{n}^{max} = 
|\vec{R}_{ij}^{max}| = 4a$ (closed symbols).
Open symbols represent the sum (using Eq.\
(\ref{fig:torque_DMI})) over all shells around site 
$i$ up to $d_{n} = |\vec{R}_{ij}|$.  Note that the antisymmetric behavior of these
interactions with respect to a permutation of sites $i$ and $j$ ensure a zero 
total torque on the magnetization due to the vanishing effective field
obtained via summation over all sites $i$ in the lattice, assuming fixed
positions of the displaced atoms.
In a corresponding manner, $|\vec{R}_{ij}|$ dependent results for the effective field due to the
diagonal anisotropic part of the SLC tensor,
i.e. $\frac{1}{2}\sum_j({\cal J}_{ij,k}^{{{x}}{{x}},\mu} - {\cal J}_{ij,k}^{{{z}}{{z}},\mu})\mbox{sin}2\theta$, are
shown in Fig.\ \ref{fig:TORQUE_diag-Jij-uj_vs_Rij}. However, their
absolute value is about an order of magnitude smaller when compared to the
effective field associated with the DMI-like SLC.

\begin{figure}[h]
\includegraphics[width=0.15\textwidth,angle=0,clip]{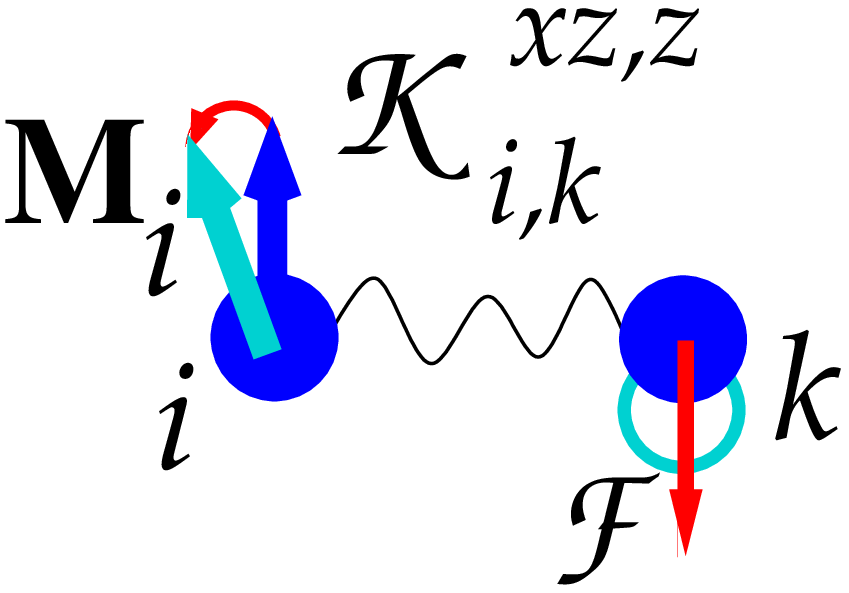}\,\,\,
\includegraphics[width=0.22\textwidth,angle=0,clip]{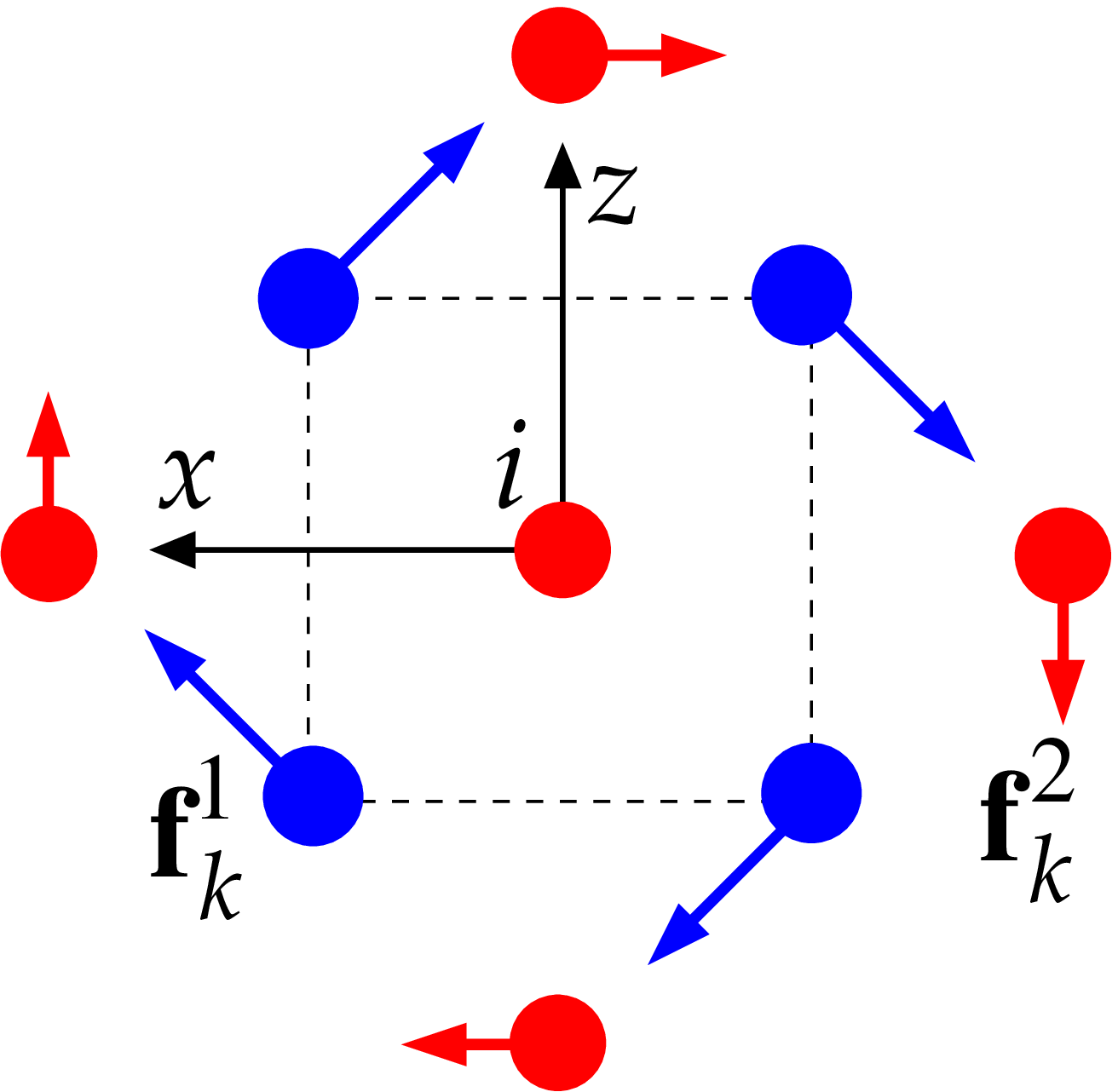}\,
\caption{\label{fig:Fe-torque_force} 
  The quantities $\vec{f} = - \sum_\mu {\cal K}^{\alpha z,\mu}_{i,k}
  \hat{n}_\mu$ ($\hat{n}_x = \hat{x}$,  $\hat{n}_y =\hat{y}$, $\hat{n}_z = \hat{z}$)
  characterizing forces on atoms
  $j$ induced by spin tilting on site $i$ via the spin-lattice 
  coupling in bcc Fe with the magnetization along $z$ axis.
  Left panel shows schematically the force $\vec{\cal F}$ on the atom
$k$ (which may be non-magnetic) induced via ${\cal K}^{\alpha
  z,\mu}_{i,k}$ SLC by tilting of spin moments on site $i$, and vice
versa, the spin tiltings induced due to 
the displacements of atoms on the site $k$.  
  Blue and red arrows show the directions of the forces on
  atoms $\vec{f}^1_k$ and $\vec{f}^2_k$ within the first and second
  atomic shells, respectively, assuming spin tilting within the $x-z$
  plane, where $\vec{f}^1_k = (\pm 0.69,\pm 0.02, \pm 0.71)$ (first shell)
  and $\vec{f}^2_k = 0.25(0,0,\pm 1)$,
  $\vec{f}^2_k = 0.25(0,\pm 1,0)$, $\vec{f}^2_k = 0.25(\pm 1,0,0)$
  (second shell).   }     
\end{figure}

Following the discussions above on the forces induced via the SLC
parameters, ${\cal J}^{{\rm dia-s},\mu}_{ij,j}$ and ${\cal
  D}^{\alpha,\mu}_{ij,j}$, one can consider also the force on the atoms
on sites $k$ induced via the MCA-like SLC ${\cal K}^{\alpha
  z,\mu}_{i,k}$ by tilting the magnetic moment on site $i$ from the
magnetization orientation $\hat{m} || \hat{z}$, as shown schematically in
Fig.\ \ref {fig:Fe-torque_force} (left panel).
The right panel represents the quantities $\vec{f} = - \sum_\mu {\cal
  K}^{\alpha z,\mu}_{i,k} \hat{n}_\mu$ characterizing forces (see Eq.\
(\ref{Eq_Torque_lattice})) on atoms $k$ (corresponding to
nearest-neighbor and next-nearest-neighbor atomic shells) induced by a 
spin tilting on site $i$ via spin-lattice coupling in bcc Fe with the
magnetization along $z$ axis. As one can see, these forces are
perpendicular to the directions connecting interacting atoms,
$\hat{R}_{ik}$, similar to the properties of the forces induced
via the DMI-like SLC shown in Fig.\ \ref{fig:Fe-Jij-Dij_force}.

One can consider a more complex example with the torque on the magnetic
moment generated by phonon-like lattice distortions instead of a single atom displacement.
As a reference state, let's consider a FM configuration in the equilibrium,
that implies zero total torque on each magnetic moment.
Creating a phonon in the system, or, e.g., switching on an external source
for acoustic waves, one can expect a distortion in the magnetic structure
induced by spin-lattice interactions.
This implies, that each spin in the FM ordered system can experience
a corresponding torque as a result of the common impact of the displaced
surrounding atoms. 
If the displacement $\vec{u}_j$ is represented in terms
of a single phonon mode $\sim e^{i(\vec(q)\cdot\vec{R}_j)}$,
the effective field calculated using this Hamiltonian is given by
\begin{eqnarray}
 H_{\rm{eff},\vec{q}}^{\alpha} &=& -\frac{\partial}{\partial e_i^{\alpha}} {\cal H}_{\rm{me-MA},\vec{q}} 
=  - \sum_{\mu} K_{i,\vec{q}}^{\alpha z,\mu}  e_i^z 
                u^{\mu}_{\vec{q}}  \;.
\label{eq:Torque_21}
\end{eqnarray}
The other way around, the SLC parameter $K_{i,\vec{q}}^{\alpha z,\mu}$ may
be seen as a force acting on the atom on site $i$ when a
periodic spin modulation occurs in the FM ordered system.
This way one can see a mutual impact of spin and lattice excitations
which can result in a simultaneous distortion in the system.

\section {Summary}

To summarize, we presented in this work a scheme to calculate the
spin-lattice coupling parameters within the multiple scattering
formalism making use of the magnetic force theorem.
The properties of the three- and four-site SLC parameters,
giving access to the SSC corrections linear and quadratic with respect
to displacements, respectively, are discussed. 
It is demonstrated that the force originating from the DMI-like SLC
parameters may be responsible for the mechanical torque on the lattice
dependent on the magnetic configuration, that can control the angular
momentum transfer via magnon-phonon scattering events.
We discussed an approach to calculate the site-diagonal SLC
parameters characterizing local magnetic anisotropy induced by a lattice
distortion, which is a counterpart to the approach based on magnetic
torque calculations worked out for the investigations of the MCA.
The approach gives access to all contributions to the MCA-like SLC
parameters, accounting also those originating from the anisotropic part of
the interatomic SLC parameters.
Furthermore, we have demonstrated the contributions
of different MCA-like SLC parameters to the energy, considering
different types of displacements.

 \appendix

\section {Computational details} \label{App:Comp_det}

The results presented in the manuscript are based on 
first-principles electronic structure calculations using the spin-polarized
relativistic Korringa Kohn Rostoker  Green function (SPR-KKR-GF) method
\cite{SPR-KKR8.5,EKM11} in combination with atomic sphere approximation (ASA).
The local spin density approximation (LSDA) to spin density
functional theory (SDFT) has been used with a parametrization for th
exchange and correlation potential as given by Vosko {\em et  al.}
\cite{VWN80}. The angular momentum expansion of the Green function was
given up to the cutoff $l_{\rm max} = 3$ was used.
A k-mesh with $36 \times 36 \times 36$ grid points
was used for the integration over the BZ.

\section {Multiple scatering formalism} \label{App:MST}

Within the KKR Green function formalism the electronic Green function 
$G(\vec{r},\vec{r}\,',E)$ is represented in real space by the expression
\cite{EBKM16}: 
\begin{eqnarray}
G(\vec{r},\vec{r}\,',E) & = &
\sum_{\Lambda_1\Lambda_2} 
Z^{n}_{\Lambda_1}(\vec{r},E)
                              {\tau}^{n n'}_{\Lambda_1\Lambda_2}(E)
Z^{n' \times}_{\Lambda_2}(\vec{r}\,',E)
 \nonumber \\
 & & 
-  \sum_{\Lambda_1} \Big[ 
Z^{n}_{\Lambda_1}(\vec{r},E) J^{n \times}_{\Lambda_1}(\vec{r}\,',E)
\Theta(r'-r)  \nonumber 
\\
 & & \qquad\quad 
J^{n}_{\Lambda_1}(\vec{r},E) Z^{n \times}_{\Lambda_1}(\vec{r}\,',E) \Theta(r-r')
\Big] \delta_{nn'} \; . \nonumber \\
\label{Eq_KKR-GF}
\end{eqnarray}
Here $\vec{r},\vec{r}'$ refer to site $n$ and $n'$, respectively,
${\tau}^{nn'}_{ \Lambda  \Lambda'} (E)$ is the so-called scattering path
operator that transfers an electronic wave coming in at 
site $ n' $ into a wave going out from site $ n $ with
all possible intermediate scattering events accounted for. 
The four-component wave functions $Z^{n}_{\Lambda}(\vec r,E)$ 
($J^{n}_{\Lambda}(\vec r,E)$) are regular (irregular)
solutions to the single-site Dirac equation with the Hamiltonian set up
within the framework of relativistic spin-density functional theory \cite{MV79,ED11}:
\begin{eqnarray}
{\cal H}_{\rm D}  & =&
- i c \vec{\bm \alpha} \cdot \vec \nabla  + \frac{1}{2} \, c^{2} ({\bm\beta} - 1) + V(\vec r)  + \beta \vec{\bm\sigma}\cdot {\vec B}_{xc}(\vec r)
\; . \nonumber \\
\label{Hamiltonian}           
\end{eqnarray}
These functions are labeled by the combined quantum numbers $\Lambda$
($\Lambda = (\kappa,\mu)$), with 
$\kappa$ and $\mu$  being the spin-orbit and magnetic quantum numbers
\cite{Ros61}. The superscript $\times$ indicates the left hand
side solution of the Dirac equation.
The operators  $ {\alpha}_i $ and $ \beta $ in the Hamiltonian in Eq.\
(\ref{Hamiltonian}) are the standard Dirac matrices \cite{Ros61}  while
 $\bar V(\vec {r}) $ and ${\vec B}_{xc}(\vec r)$ are the spin
 independent and dependent parts of the electronic potential
 \cite{Ros61,EBKM16}.

\section {Change of the inverse scattering matrix}  \label{App:UtU+}

The change of the inverse scattering matrix due to a spin tilting can be
calculated as described earlier in Ref.\ \onlinecite{EM09a},
giving this way direct access to the
derivatives w.r.t.\ $\hat{e}_i^\mu$.
In this case the change of the inverse scattering matrix $\Delta^s_{\alpha}
{\underline{m}}_i$ (the underline denotes a matrix in an
spin-angular momentum representation $\Lambda$) caused by a tilting of
spin moment on site $i$, $\delta 
\hat{e}^\alpha_i$, can be written as follows \cite{EM09a}:
\begin{eqnarray}
\Delta^s_{\alpha} {\underline{m}}_i &=& \delta \hat{e}^\alpha_i\, \underline{T}^\alpha_i\;,  \label{Eq:ME-T-1}
\end{eqnarray}
where the matrix elements $T^{\alpha}_{i,\Lambda\Lambda'}$ of the torque operator
 are given by the expression
\begin{eqnarray}
 T^{\mu}_{i,\Lambda\Lambda'} & = & \int_{\Omega_i} d^3r  \, Z^{i \times}_{\Lambda}(\vec{r},E)\, \Big[\beta \sigma_{\alpha} B_{xc}^i(\vec{r})\Big] \, Z^{
i}_{\Lambda'}(\vec{r},E)\;.  \label{Eq:ME-T-2}
\end{eqnarray}
with $\mu = (x,y,z)$.
Note that using the fixed frame of reference with the magnetization along $z$
axis, only the two torque components  $\underline{T}^{x}_i$ and
$\underline{T}^{y}_i$ linear with respect to the tilting angle are 
available. To get access to the other component $\underline{T}^{z}_i$, one has
to use a rotated frame of reference as it was suggested by Udvardi et
al. \cite{USPW03} when introducing relativistic calculations of the
exchange coupling tensor  $\underline{\underline{J}}_{ij}$ .

  
In the case of atom $i$ displaced from the equilibrium position by
$\vec{u}_i$, the change of the single-site scattering matrix $\Delta \underline{t}_i
= \underline{t}_i -  \underline{t}^0_i$ is given in terms of
the $t$-matrix $\underline{t}_i^{(0)}$ for the un-shifted atom, and the
scattering matrix
for shifted atom
\begin{eqnarray}
\underline{t}_i &=&
                    \underline{U}(\vec{u}_i)\,\underline{t}^0_i\,\underline{U}(\vec{u}_i)^{-1} \;,  
\end{eqnarray}
(analogously for the inversed scattering matrices $\underline{m}^0_i$ and $\underline{m}_i$), where
the transformation matrix   $\underline{U}_i$ is given by the expression \cite{SBZD87, PZDS97}
\begin{eqnarray}
U_{LL'}(\vec{u}_i)& = &
4\pi
                        \sum_{L''}i^{l+l''-l'}C_{LL'L''}j_{l''}(|\vec{u}_i|k){\cal
                        Y}_{L''}(\hat{u}_i) \,, \nonumber \\
\label{Eq:U-Transform}
\end{eqnarray}
given here in the non-relativistic form
with $k = \sqrt{2mE/\hbar^2}$, and ${\cal Y}_L$ real spherical harmonics.
In Eq.\ (\ref{Eq:U-Transform})
$j_{l}$ is a spherical Bessel function,
$C_{LL'L''}$  stands for the Gaunt coefficients given in
non-relativistic angular momentum representation with $L = (l,m_l)$.
The relativistic form of $U_{k,\Lambda\Lambda'}$ is obtained by a 
standard Clebsch-Gordan transformation. 
The inversed transformation matrix can be written as follows
\begin{eqnarray}
\bigg[U^{-1}(\vec{u}_i)\bigg]_{LL'}& = &U_{LL'}(-\vec{u}_i) = U_{L'L}(\vec{u}_i)\,.
\end{eqnarray}
The Bessel function $j_{l''}(|\vec{u}_i|k)$ in the limit of a small
displacement amplitude $|\vec{u}_i|$  is given by the expression
\cite{Ros57} 
\begin{eqnarray}
j_{l}(|\vec{u}_i|k) & = & \frac{(|\vec{u}_i|k)^l}{(2l+1)!!} \,.
\end{eqnarray}

\begin{widetext}
Keeping in Eq.\ (\ref{Eq:U-Transform}) only the terms up to second order
w.r.t. the displacement,  one obtains
\begin{eqnarray*}
U_{LL'}(\vec{u}_i)& = &
4\pi \sum_{L''}i^{l+l''-l'}C_{LL'L''}  
\frac{(|\vec{u}_i|k)^{l''}}{(2l''+1)!!} {\cal Y}_{L''}(\hat{u}_i) 
%
 =   
4\pi \bigg[ i^{l+0-l'}C_{LL'00}  \frac{(|\vec{u}_i|k)^0}{(1)!!} {\cal Y}_{00}(\hat{u}_i)\\
&  & +
\sum_{m=-1}^1 i^{l+1-l'}C_{LL'1m}  
\frac{(|\vec{u}_i|k)^1}{(3)!!} {\cal Y}_{1m}(\hat{u}_i) 
 +
\sum_{m=-2}^2 i^{l+2-l'}C_{LL'2m}  
\frac{(|\vec{u}_i|k)^2}{(5)!!} {\cal Y}_{2m}(\hat{u}_i) + ...\bigg]\\
%
& = &  
4\pi \bigg[ i^{l-l'} \frac{1}{\sqrt{4\pi}} 1 \frac{1}{\sqrt{4\pi}} \delta_{LL'}
   +
i^{l+1-l'} \sum_{m=-1}^1 C_{LL'1m}  \frac{|\vec{u}_i|k}{(3)} {\cal Y}_{1m}(\hat{u}_i)\\
&  &+
     i^{l+2-l'} \sum_{m=-2}^2 C_{LL'2m}  \frac{|\vec{u}_i|^2k^2}{15} C^{-1}_{l0,l'0,20} 
 \bigg(
     \sqrt{\frac{20\pi}{9}}
     \sum_{m_1=-1}^1 \sum_{m_2=-1}^1 C_{1m_1,1m_2,2m}  {\cal Y}_{1m_1}(\hat{u}_i) {\cal Y}_{1m_2}(\hat{u}_i) \bigg) + ...\bigg]\\
%
& \approx &   \delta_{LL'} + 
\frac{4\pi}{3}|\vec{u}_i|ki^{l+1-l'}\sum_{m=-1}^1 C_{LL'1m}
            {\cal Y}_{1m}(\hat{u}_i)\\
&  &+
     \frac{4\pi}{15}\sqrt{\frac{20\pi}{9}} |\vec{u}_i|^2k^2 i^{l+2-l'} \sum_{m=-2}^2 C_{LL'2m}  C^{-1}_{l0,l'0,20} 
 \bigg(    
     \sum_{m_1=-1}^1 \sum_{m_2=-1}^1 C_{1m_1,1m_2,2m} {\cal Y}_{1m_1}(\hat{u}_i)  {\cal Y}_{1m_2}(\hat{u}_i) \bigg) + ...
%
\end{eqnarray*}
where the following expansion is used \cite{Ros57}
\begin{eqnarray*}
 C_{l0,l'0,20} {\cal Y}_{lm}(\hat{u}_i) &= &  
     \sqrt{\frac{4\pi(2l+1)}{(2l_1+1)(2l_2+1))}}
     \sum_{m_1=-l_1}^{l_1} \sum_{m_2=-l_2}^{l_2} C_{l_1m_1,l_2m_2,2m} {\cal Y}_{lm_1}(\hat{u}_i) {\cal Y}_{1m_2}(\hat{u}_i)
\end{eqnarray*}
Here the direction of displacenent of atom $i$ is given by unit vector $\hat{u}_i$. 

Representing the real spherical harmonics ${\cal Y}_{1m}(\hat{u}_i)$ in the following form
\begin{eqnarray*}
{\cal Y}_{1m}(\hat{u}_i)& = &\sqrt{\frac{3}{4\pi}}
\left\{
\begin{array}{ccc}
u_i^y/|\vec{u}_i| = \hat{u}_i^y & {\rm for} & m=-1 \\
u_i^z/|\vec{u}_i| = \hat{u}_i^z & {\rm for} & m=0 \\
u_i^x/|\vec{u}_i| = \hat{u}_i^x & {\rm for} & m=+1
\end{array} \right. 
\end{eqnarray*}


the transformation functions $U_{LL'}(\vec{u}_i)$ is reduced to the following form
\begin{eqnarray*}
U_{LL'}(\vec{u}_i)
& \approx &   \delta_{LL'} + 
 \sum_{\mu} \frac{4\pi}{3}\sqrt{\frac{3}{4\pi}} {u}_i^\mu ki^{l+1-l'} C_{LL'1m_\mu}   \\
&  &+
    \sum_{\mu} \sum_{\nu}^1 \frac{4\pi}{15}\sqrt{\frac{20\pi}{9}}\frac{3\cdot2}{4\pi}
      {u}_i^{\mu}{u}_i^{\nu} k^2 i^{l+2-l'} \sum_{m=-2}^2 C_{LL'2m}  C^{-1}_{l0,l'0,20} 
     C_{1m_\mu,1m_\nu,2m}  + ... \\
& = & \delta_{LL'} + 
 \sum_{\mu} {u}_i^\mu \bar{U}^{m_\mu}_{LL'} +
      \sum_{\mu\nu} {u}_i^\mu{u}_i^\nu\bar{U}^{(2b),m_\mu m_\nu}_{LL'}\\  
  U_{LL'}(-\vec{u}_i) & \approx & \delta_{LL'} -
 \sum_{\mu} {u}_i^\mu \bar{U}^{m_\mu}_{LL'} +
      \sum_{\mu\nu} {u}_i^\mu{u}_i^\nu\bar{U}^{(2b),m_\mu m_\nu}_{LL'}                                  
\end{eqnarray*}
where $m_{\mu(\nu)} = \{m_{x}, m_{y} m_{z}\}$, such that $m_{x} = 1,\,m_{y} = -1,\, m_{z} = 0$,



\begin{eqnarray*}
\bar{U}_{LL'}^{m_x} &=& k \frac{4\pi}{3}i^{l-l'+1}\sqrt{\frac{3}{4\pi}} C_{LL'1\;+1} 
= ki^{l-l'+1} \sqrt{\frac{4\pi}{3}} C_{LL'1\; +1} \\
\bar{U}_{LL'}^{m_y} &=& k i^{l-l'+1}\sqrt{\frac{4\pi}{3}} C_{LL'1\; -1} \\
\bar{U}_{LL'}^{m_z} &=& k i^{l-l'+1}\sqrt{\frac{4\pi}{3}} C_{LL'1\; 0} \\
  \bar{U}^{2b,\mu \nu}_{LL'} & \equiv &\bar{U}^{2b,m_\mu m_\nu}_{LL'} =  k^2 \frac{1}{3}\sqrt{\frac{4\pi}{5}} i^{l+2-l'} \sum_{m=-2}^2 C_{LL'2m}  C^{-1}_{l0,l'0,20}    
      C_{1m_\mu,1m_\nu,2m} 
\end{eqnarray*}
where
\begin{eqnarray*}
 \bar{U}^\mu_{LL'}(-\hat{u}_i) &=& -\bar{U}^\mu_{LL'}(\hat{u}_i) \\
  \bar{U}^{2b,\mu \nu}_{LL'}(-\hat{u}_i) &=& \bar{U}^{2b,\mu \nu}_{LL'}(\hat{u}_i)
\end{eqnarray*}
as a consequence of the property
\begin{eqnarray*}
{\cal Y}_{1m}(-\hat{u}_i)& = & (-1)^1 {\cal Y}_{1m}(\hat{u}_i) = -{\cal Y}_{1m}(\hat{u}_i)
\end{eqnarray*}



Thus, we obtain the approximate transformation matrix for a small displacement $\vec{u}^{\,\mu}_i$
\begin{eqnarray}
U_{LL'}(\vec{u}_i) & = &  \delta_{LL'} +     \sum_{\mu}  u^{\mu}_i   \,
                         \bar{U}_{LL'}^{m_\mu} +\sum_{\mu, \nu}
                         {u}_i^\mu{u}_i^\nu \bar{U}^{(2b),m_\mu
                         m_\nu}_{LL'}\,,                         
\end{eqnarray}
or, using the notation $\bar{U}^{(2b),\mu\nu}_{LL'} \equiv \bar{U}^{(2b),m_\mu
  m_\nu}_{LL'}$, one can represent the modified single-site scattering
matrix as follows
\begin{eqnarray*}
\tilde{\underline{t}}_i& = & \underline{U}(\vec{u}_i)\,\underline{t}_i\,\underline{U}^{-1}(\vec{u}_i) 
         =  \underline{U}(\vec{u}_i)\,\underline{t}_i\,\underline{U}(-\vec{u}_i)     \\
           & \approx & \bigg( \underline{I} +   \sum_{\mu'} u^{\mu'}_i   \,
                        \underline{\bar{U}}^{\mu'}(\hat{u}_{i}) +
                       \sum_{\mu'\nu'}  {u}_i^{\mu'}{u}_i^{\nu'}
                       \underline{\bar{U}}^{(2b),\mu'\nu'}(\hat{u}_i) \bigg)\,
                       \underline{t}_i \, 
            \bigg( \underline{I} +  \sum_{\nu"} u^{\nu"}_i   \,
                        \underline{\bar{U}}^{\mu"}(-\hat{u}_{i}) +
                      \sum_{\mu"\nu"} {u}_i^{\mu"}{u}_i^{\nu"}
                      \underline{\bar{U}}^{(2b),\mu"
                         \nu"}(-\hat{u}_i) \bigg) \\
         & \approx & \underline{t} + \sum_{\mu} \bigg( u^{\mu}_i \, \bar{U}^{\,\mu}(\hat{u}_i) \, \underline{t}_i + u^{\mu}_i \, \underline{t}_i \, \bar{U}^{\,\mu}(-\hat{u}_i) \bigg)\\ 
 && + \sum_{\mu\nu} u^{\mu}_i u^{\nu}_i \bigg( \, \bar{U}^{\mu}(\hat{u}_i) \,\underline{t}_i \,
   \bar{U}^{\,\nu}(-\hat{u}_i) +  \bar{U}(\hat{u}^{\nu}_i) \,\underline{t}_i \,
   \bar{U}(-\hat{u}^{\,\mu}_i) 
            + \underline{\bar{U}}^{(2b),\mu
                         \nu}(\hat{u}_i)\underline{t}_i
            + \underline{t}_i \,\underline{\bar{U}}^{(2b),\mu
                         \nu}(-\hat{u}_i)\bigg)
\end{eqnarray*}

\begin{eqnarray}
 \tilde{\underline{t}}_i -  \underline{t}_i& = & \sum_{\mu}  u^{\mu}_i
                                                 \,\bigg(\bar{U}^{\,\mu}(\hat{u}_i)\,\underline{t}_i
                                                 +
                                                 t_i\,\bar{U}^{\,\mu}(-\hat{u}_i)\bigg)
  \nonumber \\
                                           & &+ \sum_{\mu\nu} u^{\mu}_i u^{\nu}_i  \,
\bigg( \, \bar{U}^{\mu}(\hat{u}_i) \,\underline{t}_i \,
   \bar{U}^{\,\nu}(-\hat{u}_i) + \bar{U}^{\nu}(\hat{u}_i) \,\underline{t}_i \,
   \bar{U}^{\,\mu}(-\hat{u}_i)
            + \underline{\bar{U}}^{(2b),\mu
                         \nu}(\hat{u}_i)\underline{t}_i
            + \underline{t}_i \,\underline{\bar{U}}^{(2b),\mu
                         \nu}(-\hat{u}_i)\bigg)
                                               +O\bigg((u^{\mu}_i)^3
                                               \bigg)\nonumber \\
& = & \sum_{\mu}  u^{\mu}_i
      \,\bigg(\bar{U}^{\,\mu}(\hat{u}_i)\,\underline{t}_i -
      \underline{t}_i\,\bar{U}^{\,\mu}(\hat{u}_i)\bigg) \nonumber \\
                                           & &+ \sum_{\mu\nu} u^{\mu}_i u^{\nu}_i  \,
\bigg( \,- \bar{U}^{\,\mu}(\hat{u}_i) \,\underline{t}_i \,
   \bar{U}^{\,\nu}(\hat{u}_i) -  \bar{U}^{\,\nu}(\hat{u}_i) \,\underline{t}_i \,
   \bar{U}^{\,\mu}(\hat{u}_i)
            + \underline{\bar{U}}^{(2b),\mu
                         \nu}(\hat{u}_i)\underline{t}_i
            + \underline{t}_i \,\underline{\bar{U}}^{(2b),\mu
                         \nu}(\hat{u}_i)\bigg)
                                               +O\bigg((u^{\mu}_i)^3 \bigg)\,,  
\end{eqnarray}

and analogously for $m_i=t^{-1}_i$.


\end{widetext}

\bigskip

\section{Torque on magnetic moment} \label{App:Torque}

Here we give the relationship between the torque on a magnetic
moment of the system and the energy change due to a rotation of the magnetic
moment, that is used for the calculations of the magneto-crystalline
anisotropy energy in magnetic systems \cite{SSB+06}.
Let us consider a FM-ordered system with the magnetization direction
$\hat{e}$. 
The energy change due to a tilting of the magnetic moment is given by the expression
\begin{eqnarray}
\delta E   &=&  \frac{\delta E}{\delta \hat{e}} \cdot \delta \hat{e} =
               \frac{\delta E}{\delta \hat{e}} \cdot \vec{\delta \theta} \times
               \hat{e} \\  
         &=& - \vec{H}_{eff} \cdot [\vec{\delta \theta} \times
               \hat{e}] = - \vec{\delta \theta} \cdot [\hat{e} \times
             \vec{H}_{eff}]   \\  
         &=&    - \delta \theta \,\, \hat{n} \cdot [\hat{e} \times
             \vec{B}_{eff}] = \delta  \theta \, T^{\hat{n}}
\label{eq:Totque_3}        
\end{eqnarray}
with the effective field $\vec{H}_{eff} = -\frac{\delta E}{\delta \hat{e}}$, $\delta \hat{e} =
\vec{\delta \theta} \times \hat{e} $, $\vec{\delta \theta} = \hat{n}
\delta \theta $, and  ${\hat{n}}$ the direction perpendicular to 
 the plane of rotation by the angle $\delta \theta$ of the magnetization direction.
Thus, the torque on the magnetic moment represented in terms of local
effective field 
\begin{eqnarray}
\vec{T}  &=& \hat{e} \times \vec{H}_{eff}             
\end{eqnarray}
gives access to the MCA energy via its projection on the direction ${\hat{n}}$
\begin{eqnarray}
T^{\hat{n}}(\hat{e})   &=& \hat{n} \cdot [\hat{e} \times
             \vec{H}_{eff}]             
\end{eqnarray}
characterising the energy change due to a rotation of the magnetization.
Representing this direction in terms of the polar angles $\theta$
and $\phi$, the torque may be defined as the derivative
\begin{eqnarray}
T^{\hat{n}}(\hat{e})  &=&   - \frac{\partial  E}{\partial \theta}  \,.    
\end{eqnarray}
As it was discussed in  Refs.\ \cite{SSB+06,USPW03}, this quantity can be
used for the calculation of the magnetic anisotropy parameters. In
particular,  considering the magnetization direction tilted by $\theta =
\pi/4$, the corresponding torque $T^{\hat{n}}(\pi/4)$ gives  direct
access to the energy of uniaxial anisoropy $T^{\hat{n}}(\pi/4) =
E(\hat{e}||\hat{z}) - E(\hat{e}||\hat{x})$, and as a consequence to the
uniaxial anisoropy parameters.


%

\end{document}